\documentclass[superscriptaddress,amsmath,amssymb,aps,prd,preprintnumbers,showpacs,twocolumn,nofootinbib]{revtex4-2}

\usepackage[T1]{fontenc}
\usepackage[utf8]{inputenc}
\usepackage{amsmath,amssymb,bm,mathtools,natbib}
\usepackage{graphicx}
\usepackage[caption=false]{subfig}
\usepackage{slashed}
\usepackage{dsfont}
\usepackage{float}
\usepackage{cancel}
\usepackage{units}
\usepackage{array}
\usepackage{upgreek}
\usepackage{booktabs}
\usepackage[dvipsnames,table,xcdraw]{xcolor}
\usepackage{enumerate}
\usepackage[normalem]{ulem}
\usepackage[colorlinks=true,pdfstartview=FitV,linkcolor=blue,citecolor=blue,urlcolor=blue,breaklinks=true]{hyperref}
\usepackage{orcidlink}

\begin{document}

\title{Geometric Phases and Persistent Spin Currents \\ from nonminimal couplings}

\author{Jo\~{a}o A.A.S.\ Reis\orcidlink{0000-0002-2831-5317}}
\email{joao.reis@uesb.edu.br}
\affiliation{Departamento de Ci\^{e}ncias Exatas e Naturais, \\ Universidade Estadual do Sudoeste da Bahia, Itapetinga (BA), 45700-000, Brazil}

\author{L. Lisboa-Santos\orcidlink{0000-0003-4939-3856}}
\email{leticia.lisboa@discente.ufma.br}
\affiliation{Programa de Pós-Graduação em Física, Universidade Federal do Maranh\~{a}o, Campus Universit\'{a}rio do Bacanga, S\~{a}o Lu\'{i}s (MA), 65085-580, Brazil}

\author{Edilberto O. Silva\orcidlink{0000-0002-0297-5747}}
\email{edilberto.silva@ufma.br}
\affiliation{Departamento de F\'{\i}sica, Universidade Federal do Maranh\~{a}o, 65085-580 S\~{a}o Lu\'{\i}s, Maranh\~{a}o, Brazil}

\date{\today}

\begin{abstract}
We investigate a class of nonminimal derivative couplings between fermions
and electromagnetic fields that generate Rashba-like spin--orbit interactions
in one-dimensional quantum rings. Starting from a generalized Dirac Lagrangian
containing two independent axial structures built from the field strength
$F_{\mu\nu}$ and its dual $\tilde{F}_{\mu\nu}$, we perform a systematic
nonrelativistic expansion and show that both couplings induce effective
Hamiltonians of the form $\boldsymbol{\mathcal{F}}\cdot(\boldsymbol{p}\times
\boldsymbol{\sigma})$. This reveals that magnetic as well as electric
background fields may give rise to Rashba-type interactions, in contrast
with standard condensed-matter scenarios. Before passing to the
nonrelativistic limit, we analyze the relativistic content of the model
in detail: the canonical structure of the deformed Dirac operator, the
admissible background classes, the effective bilinear current, and the
branch splitting of the relativistic dispersion relation, which constitutes
the primary relativistic signature of the theory. We derive exact analytical
energy levels and normalized eigenspinors for the resulting ring Hamiltonian,
compute Aharonov--Anandan geometric phases, and analyze persistent spin
currents together with the associated differential spin response
$\mathcal{G}_s = \partial\mathcal{J}_\varphi^z/\partial\xi$. Exploiting the analytical control offered by the
model, we derive the first systematic order-of-magnitude bounds on the two
Lorentz-invariant couplings $\mathfrak{g}_1$ and $\mathfrak{g}_2$ from
both spectroscopic and mesoscopic scenarios, identifying the experimental
channels most sensitive to the new physics encoded in these operators.
We discuss physical implications, signatures, and possible experimental
analogs, and outline several promising directions involving disorder,
noise, and nonequilibrium spin dynamics.
\end{abstract}

\maketitle

%\tableofcontents

\section{Introduction}

Spin--orbit coupling constitutes one of the central organizing principles
of modern condensed-matter and mesoscopic physics, governing phenomena
such as the spin Hall effect~\cite{Murakami2003,Sinova2004},
persistent spin currents, geometric phases, and the emergence of
topological states of matter~\cite{Kane2005,Bernevig2006,Hasan2010}.
Among the various realizations of spin--orbit coupling, the Rashba
mechanism~\cite{Rashba1960,Bychkov1984} occupies a privileged position:
it arises whenever structural inversion symmetry is broken, it is
intrinsically tunable through external gate voltages or electromagnetic
fields~\cite{Nitta1997}, and it directly links spin dynamics to orbital
motion in a geometrically transparent way. Since its original formulation,
the Rashba interaction has become a cornerstone of
spintronics~\cite{Zutic2004,Manchon2015}, providing the microscopic
foundation for spin-field-effect transistors~\cite{Datta1990},
spin-orbit torques~\cite{Miron2011}, and topological
edge states~\cite{Kane2005,Bernevig2006}.

A complementary thread in mesoscopic physics concerns the geometric
structure of quantum states in parameter space. Following the seminal work
of Berry~\cite{Berry1984} on adiabatic phases, Aharonov and
Anandan~\cite{AharonovAnandan1987} generalized the concept to arbitrary
cyclic evolutions, showing that any closed path in Hilbert space
accumulates a gauge-invariant geometric phase. In mesoscopic systems
threaded by magnetic flux, the Aharonov--Bohm effect~\cite{AharonovBohm1959}
provides the prototypical realization of such phase sensitivity.
When spin--orbit coupling is present, the interplay between the
Aharonov--Bohm and Aharonov--Anandan phases gives rise to nontrivial
interference patterns~\cite{Frustaglia2004,Bercioux2005} whose
experimental detection in semiconductor rings has been
reported~\cite{Nitta1997,Morpurgo1998,Konig2006}.

Quantum rings constitute especially fertile platforms for exploring
these phenomena~\cite{Lorke2000,Fuhrer2001}. Their one-dimensional
topology forces the spinor to complete a closed trajectory at each
revolution, converting the spatial periodicity into a direct probe of
phase accumulation and spin precession. The persistent charge
current~\cite{Buttiker1983,Cheung1988,Levy1990,Chandrasekhar1991}
and, more recently, persistent spin
currents~\cite{Malshukov2002,Sun2004,Frustaglia2004,Sheng2006} have
been studied extensively in this geometry. The sensitivity of these
quantities to the Rashba coupling parameter makes quantum rings
natural detectors of any mechanism that modifies the effective
spin--orbit structure of the one-particle Hamiltonian.

From a fundamental perspective, the question of how spin--orbit
interactions can emerge from relativistic quantum field theory---rather
than being postulated phenomenologically---motivates the exploration of
nonminimal couplings in the Dirac equation. Such couplings arise naturally
in several distinct theoretical contexts. In the Standard Model
Extension (SME)~\cite{Colladay1997,Colladay1998,Kostelecky2004},
nonminimal operators encode departures from Lorentz and CPT symmetry,
and their fermion--photon sector has been systematically
classified~\cite{PhysRevD.99.056016}. In axion electrodynamics and
axion-like particle models~\cite{Wilczek1987,Sikivie1983}, derivative
couplings between fermions and electromagnetic backgrounds appear
as effective operators generated by integrating out the axion field.
In emergent gauge theories of condensed-matter
systems~\cite{Volovik2003,Ryu2010}, analogous structures arise as
low-energy descriptions of topological phases. In each of these
frameworks, the nonminimal coupling enters as a dimension-six operator
suppressed by a high-energy scale, and its physical content is carried
by the combination of the Wilson coefficient with the background
electromagnetic field.

In this work we revisit a general Dirac Lagrangian containing two
independent nonminimal derivative couplings: one involving $F_{\mu\nu}$
and another involving the dual tensor $\tilde{F}_{\mu\nu}$, both of
the axial type. The two operators are Lorentz-invariant and belong to
the dimension-six fermion--photon sector of the general
effective-field-theory framework of Ref.~\cite{PhysRevD.99.056016}.
Although the full relativistic dynamics is involved, we show that the
corresponding nonrelativistic limit yields a simple and unified effective
Hamiltonian. Both sectors generate a term of the form
$\boldsymbol{\mathcal{F}}\cdot(\boldsymbol{p}\times\boldsymbol{\sigma})$,
implying that not only electric fields but also magnetic fields can
induce Rashba-like interactions, in sharp contrast with the standard
condensed-matter scenario where the Rashba coupling is exclusively
driven by structural electric fields~\cite{Bychkov1984}.

After deriving the effective Hamiltonian, we apply it to electrons
confined to a one-dimensional quantum ring. We obtain exact analytical
expressions for the spectrum and eigenspinors, compute associated
geometric phases of the Aharonov--Anandan type, and evaluate persistent
spin currents. We also analyze the thermodynamic behavior using both
the canonical and grand-canonical ensembles, and extend the model to
include interactions through a mean--field density-dependent potential.
This allows us to identify how the modified dispersion relation affects
thermal, magnetic and transport properties. Finally, exploiting the
analytical control offered by the model, we extract the first
systematic order-of-magnitude bounds on the couplings $\mathfrak{g}_1$
and $\mathfrak{g}_2$ from spectroscopic and mesoscopic observables.

The results reveal novel features of electromagnetic-field-induced
Rashba systems, including modified persistent currents, distinctive
phase contributions, nontrivial thermodynamic responses, and a
differential spin response $\mathcal{G}_s$ that plays the role of a
mesoscopic spin conductance. Possible connections with axion-like
models~\cite{Wilczek1987}, synthetic gauge
fields~\cite{Dalibard2011,Goldman2014}, and ring-shaped experimental
platforms are briefly discussed.

\section{General model}
\label{sec:model}

This section establishes the relativistic starting point of the analysis and identifies the background structures that later generate the effective spin--orbit interaction. Our aim here is not yet to derive the low-energy Hamiltonian, but rather to show, already at the Dirac level, why the present class of couplings is capable of producing nontrivial geometric spin dynamics.

We consider the interaction Lagrangian
\begin{equation}
\mathcal{L}_{\mathrm{int}}
=
\frac{1}{2}\,\bar{\psi}\Big(
\mathfrak{g}_{1}\,F_{\mu\nu}\gamma^{\mu}\gamma_{5}\,iD^{\nu}
+
\mathfrak{g}_{2}\,\tilde{F}_{\mu\nu}\gamma^{\mu}\gamma_{5}\,iD^{\nu}
\Big)\psi
+\mathrm{H.c.},
\label{eq:Lint-model}
\end{equation}
where
\begin{equation}
\tilde{F}_{\mu\nu}=\frac{1}{2}\epsilon_{\mu\nu\rho\sigma}F^{\rho\sigma},
\qquad
D_{\mu}=\partial_{\mu}+ieA_{\mu}.
\end{equation}
The physical content of Eq.~\eqref{eq:Lint-model} is worth emphasizing from the outset. The coupling $\mathfrak g_{1}$ probes the ordinary electromagnetic tensor, whereas $\mathfrak g_{2}$ probes its dual. More importantly, both interactions are \emph{derivative} couplings: the background does not simply act as an external potential, but rather enters the operator that controls the propagation of the fermion itself. This is the basic reason why the model is able to generate effective momentum--spin structures instead of only static energy shifts.

In order to place Eq.~\eqref{eq:Lint-model} within a broader effective-field-theory framework, it is convenient to relate it to the general dimension-six fermion--photon sector \cite{PhysRevD.99.056016}. A useful parametrization is
\begin{equation}
\mathcal{L}^{(6)}_{d_F}
=
\frac{1}{2}\,\bar{\psi}
\Big(
d_F^{(6)\mu\alpha\beta\gamma}\,
F_{\beta\gamma}\,
\gamma_{5}\gamma_{\mu}\,
iD_{\alpha}
\Big)\psi
+\mathrm{H.c.},
\label{eq:LdF-general}
\end{equation}
where $d_F^{(6)\mu\alpha\beta\gamma}$ is a rank-four tensor encoding the admissible nonminimal couplings. The particular model studied here is obtained by the Lorentz-invariant axial truncation
\begin{equation}
d_{F,\mathrm{model}}^{(6)\mu\alpha\beta\gamma}
=
\frac{\mathfrak{g}_{1}}{2}
\left(
\eta^{\mu\beta}\eta^{\alpha\gamma}
-
\eta^{\mu\gamma}\eta^{\alpha\beta}
\right)
+
\frac{\mathfrak{g}_{2}}{2}\epsilon^{\mu\alpha\beta\gamma}.
\label{eq:dF-mapping}
\end{equation}
Substituting Eq.~\eqref{eq:dF-mapping} into Eq.~\eqref{eq:LdF-general}, one obtains
\begin{align}
\frac{\mathfrak g_1}{2}
\left(
\eta^{\mu\beta}\eta^{\alpha\gamma}
-
\eta^{\mu\gamma}\eta^{\alpha\beta}
\right)F_{\beta\gamma}
&=
\mathfrak g_1 F^{\mu\alpha},
\\
\frac{\mathfrak g_2}{2}\epsilon^{\mu\alpha\beta\gamma}F_{\beta\gamma}
&=
\mathfrak g_2 \tilde F^{\mu\alpha},
\end{align}
where the antisymmetry of $F_{\beta\gamma}$ has been used in the first identity. These intermediate steps make clear that the two couplings retained in Eq.~\eqref{eq:Lint-model} correspond to two independent tensorial sectors of the same effective theory.

This EFT interpretation is useful for two reasons. First, it shows that the model is not an ad hoc ansatz, but a well-defined truncation of a broader relativistic operator basis. Second, it makes clear that the two branches, although they may later lead to similar low-energy structures, are microscopically distinct from the start: one is tied to $F_{\mu\nu}$ and the other to $\tilde F_{\mu\nu}$.

The interaction in Eq.~\eqref{eq:Lint-model} is written in the standard $\bar\psi\,\Gamma\,iD\psi+\mathrm{H.c.}$ form. An equivalent formulation can be written using the symmetrized derivative $i\overleftrightarrow{D}_{\mu}$, differing only by total derivatives and terms involving derivatives of the background fields, which are irrelevant under suitable boundary conditions. Since the operators in Eq.~\eqref{eq:Lint-model} have mass dimension six, the couplings satisfy
\begin{equation}
[\mathfrak g_1]=[\mathfrak g_2]=-2,
\end{equation}
and should therefore be interpreted as effective parameters suppressed by a high-energy scale.

Adding the free Dirac sector,
\begin{equation}
\mathcal L
=
\bar\psi(i\gamma^\mu D_\mu-m)\psi+\mathcal L_{\mathrm{int}},
\end{equation}
and varying with respect to $\bar\psi$, we obtain the modified Dirac equation
\begin{equation}
\left(
i\gamma^{\mu}D_{\mu}
+\mathfrak{g}_{1}F_{\mu\nu}\gamma^{\mu}\gamma_{5}iD^{\nu}
+\mathfrak{g}_{2}\tilde{F}_{\mu\nu}\gamma^{\mu}\gamma_{5}iD^{\nu}
-m
\right)\psi=0.
\label{eq:Dirac-full}
\end{equation}
Equation \eqref{eq:Dirac-full} is the fundamental relativistic object of the model. It already shows that the electromagnetic background affects the one-particle dynamics through the operator multiplying the derivatives, not merely through an additive interaction term. For that reason, the resulting theory is sensitive to the orientation of the background tensors and to the way in which they contract with the fermion four-momentum. This is the first indication that the low-energy sector will contain genuinely geometric spin-dependent couplings rather than only conventional Zeeman-like contributions.

It is useful to rewrite Eq.~\eqref{eq:Dirac-full} in the compact form
\begin{equation}
\left(
i\Gamma^\nu_{\mathrm{eff}}D_\nu-m
\right)\psi=0,
\label{eq:Dirac-compact-model}
\end{equation}
with
\begin{equation}
\Gamma^\nu_{\mathrm{eff}}
=
\gamma^\nu
+\mathfrak g_1 F_{\mu}{}^{\nu}\gamma^\mu\gamma_5
+\mathfrak g_2 \tilde F_{\mu}{}^{\nu}\gamma^\mu\gamma_5.
\label{eq:Gammaeff-model}
\end{equation}
This rewriting is more than a compact notation. It makes explicit that the model may be viewed as a Dirac theory in which the kinetic matrices themselves are deformed by an axial electromagnetic background. In other words, the relativistic dynamics is modified at the level of propagation, not merely at the level of external forcing.

This observation provides the natural bridge to the next section. Rather than immediately reducing Eq.~\eqref{eq:Dirac-full} to its nonrelativistic limit, we first examine the relativistic content of the effective operator in Eq.~\eqref{eq:Gammaeff-model}: its canonical structure, the conditions under which the time evolution remains well posed, the effective bilinear current associated with the deformed kinetic sector, and the spectral problem defined by plane-wave propagation in constant backgrounds. Only after this relativistic analysis do we pass to the low-energy Hamiltonian and its realization on a quantum ring.

\section{Relativistic Dirac dynamics from nonminimal axial derivative couplings}
\label{sec:relativistic}

We now take Eq.~\eqref{eq:Dirac-compact-model} as the central object of the analysis. The purpose of this section is to extract the genuinely relativistic content of the theory before passing to the Foldy--Wouthuysen reduction. This step is important for the logic of the paper. If one moves too quickly to the low-energy regime, the effective spin--orbit interaction may look like a formal byproduct of the expansion. By contrast, once the relativistic equation is examined on its own terms, one sees that the later Rashba-like structure is the infrared manifestation of a more fundamental deformation of the Dirac kinetic operator by electromagnetic tensor backgrounds.

\paragraph{Canonical structure and admissible relativistic sectors.}
Starting from
\begin{equation}
\left(
i\Gamma^\nu_{\mathrm{eff}}D_\nu-m
\right)\psi=0,
\label{eq:Dirac-compact-rel}
\end{equation}
the first question is whether the modified equation still defines a sensible first-order evolution problem. Isolating the time derivative, one finds
\begin{equation}
i\Gamma^0_{\mathrm{eff}}\partial_0\psi
=
\left[
-\,i\Gamma^j_{\mathrm{eff}}\partial_j
+ m
+ e\,\Gamma^\mu_{\mathrm{eff}}A_\mu
\right]\psi,
\label{eq:canonical-rel}
\end{equation}
with
\begin{equation}
\Gamma^0_{\mathrm{eff}}
=
\gamma^0
+\mathfrak g_1 F_{\mu 0}\gamma^\mu\gamma_5
+\mathfrak g_2 \tilde F_{\mu 0}\gamma^\mu\gamma_5.
\label{eq:Gamma0-rel}
\end{equation}

Equation \eqref{eq:Gamma0-rel} is already highly informative. It shows that the nonminimal background may deform not only the spatial propagation of the fermion, but also the coefficient of the first time derivative itself. This is the point at which the relativistic theory acquires conceptual autonomy. Before any low-energy approximation is performed, one must first identify the classes of background for which the Cauchy problem remains well posed and the one-particle Hamiltonian interpretation survives.

Whenever $\Gamma^0_{\mathrm{eff}}$ is invertible, Eq.~\eqref{eq:canonical-rel} can be written formally as
\begin{equation}
i\partial_0\psi = H_{\mathrm{eff}}\psi, \label{eq:Heff-formal-rel}
\end{equation}
with
\begin{equation}
H_{\mathrm{eff}}
=
(\Gamma^0_{\mathrm{eff}})^{-1}
\left[
-\,i\Gamma^j_{\mathrm{eff}}\partial_j
+ m
+ e\,\Gamma^\mu_{\mathrm{eff}}A_\mu
\right].
\label{eq:Heff}
\end{equation}
This expression should be read as a structural statement rather than as the final Hamiltonian formula to be used later. Its role here is to make explicit that the very existence of a Hamiltonian description hinges on the temporal operator $\Gamma^0_{\mathrm{eff}}$.

A particularly clean and physically transparent subsector is obtained by imposing
\begin{equation}
F_{i0}=0,
\qquad
\tilde F_{i0}=0.
\label{eq:static-subclasses-rel}
\end{equation}
Under these conditions the nonminimal terms do not deform the coefficient of the first time derivative, and the relativistic theory preserves the standard Dirac time-evolution structure. These static sectors therefore play a distinguished role throughout the paper: they are not merely convenient simplifications, but the simplest classes of backgrounds in which the relativistic dynamics admits a manifestly Hermitian one-particle interpretation.

More general background configurations may in principle be treated by means of a field redefinition,
\begin{equation}
\psi=\mathcal B\,\Psi,
\label{eq:field-redef-rel}
\end{equation}
chosen so as to restore a canonical temporal structure. We do not need that construction here in full generality. What matters for the present discussion is that the temporal sector itself carries physical information: the admissibility of a background is controlled not only by how it affects the spatial couplings, but also by how it enters the time derivative of the Dirac equation.

\paragraph{Adjoint equation and effective bilinear current.}
Because the derivative sector is deformed, the bilinear structure of the relativistic theory must also be reconsidered. Taking the adjoint of Eq.~\eqref{eq:Dirac-compact-rel}, one obtains
\begin{equation}
\bar\psi
\left(
-\,i\overleftarrow{D}_{\nu}\Gamma^\nu_{\mathrm{eff}}-m
\right)=0.
\label{eq:adjoint-rel}
\end{equation}
Combining Eqs.~\eqref{eq:Dirac-compact-rel} and \eqref{eq:adjoint-rel}, one is naturally led to define the effective current
\begin{equation}
J^\mu_{\mathrm{eff}}
=
\bar\psi\,\Gamma^\mu_{\mathrm{eff}}\,\psi.
\label{eq:Jeff-rel}
\end{equation}

For constant backgrounds, or more generally for slowly varying backgrounds such that derivative corrections may be neglected at the order of interest, the equations of motion imply
\begin{equation}
\partial_\mu J^\mu_{\mathrm{eff}}=0.
\label{eq:continuity-rel}
\end{equation}
This current is the natural generalization of the usual Dirac current in the present framework. The point is physically important: once the background deforms the kinetic operator, the relevant density is no longer a priori the canonical $\psi^\dagger\psi$, but rather the density induced by the effective matrices themselves.

A short intermediate manipulation makes this point transparent. Multiplying Eq.~\eqref{eq:Dirac-compact-rel} on the left by $\bar\psi$ and Eq.~\eqref{eq:adjoint-rel} on the right by $\psi$, one obtains
\begin{align}
\bar\psi\,\Gamma^\mu_{\mathrm{eff}}D_\mu\psi
&=
-i m\,\bar\psi\psi,
\\
(\overleftarrow{D}_{\mu}\bar\psi)\,\Gamma^\mu_{\mathrm{eff}}\psi
&=
+i m\,\bar\psi\psi .
\end{align}
Adding the two relations, the mass terms cancel and one is left with the divergence of the bilinear form in Eq.~\eqref{eq:Jeff-rel}, up to derivatives of the background tensors. In the constant-background sector this immediately gives Eq.~\eqref{eq:continuity-rel}.

In particular,
\begin{equation}
J^0_{\mathrm{eff}}=\bar\psi\,\Gamma^0_{\mathrm{eff}}\,\psi
\label{eq:J0eff-rel}
\end{equation}
is directly tied to the same structure that controls the well-posedness of the time evolution. Thus, the temporal sector plays a dual role: it governs both the existence of a sensible Hamiltonian formulation and the appropriate notion of relativistic probability density.

\paragraph{Discrete symmetries and physical distinction between the two branches.}
Although the two couplings retained in the model lead to closely related spin--orbit structures at low energies, they are not microscopically equivalent. The $\mathfrak g_1$ branch couples the axial fermion structure to the ordinary electromagnetic tensor $F_{\mu\nu}$, whereas the $\mathfrak g_2$ branch couples it to the dual tensor $\tilde F_{\mu\nu}$. This distinction is already meaningful at the relativistic level and can be sharpened by recalling the standard transformation properties of the background fields:
\begin{equation}
P:\qquad \mathbf E\rightarrow -\mathbf E,\qquad \mathbf B\rightarrow \mathbf B,
\label{eq:parity-backgrounds}
\end{equation}
\begin{equation}
T:\qquad \mathbf E\rightarrow \mathbf E,\qquad \mathbf B\rightarrow -\mathbf B.
\label{eq:time-backgrounds}
\end{equation}
At the same time, the spatial derivative changes sign under both $P$ and $T$, while the spatial axial structure carried by $\gamma^i\gamma_5$ behaves as a spin-like object.

These simple facts already explain why the two branches are inequivalent. In the static magnetic-type sector,
\begin{equation}
F_{ij}=\epsilon_{ijk}B^k,
\label{eq:Fij-B-rel}
\end{equation}
whereas in the static electric-type sector,
\begin{equation}
\tilde F_{ij}=-\epsilon_{ijk}E^k.
\label{eq:Fdualij-E-rel}
\end{equation}
Thus the two interactions couple the same axial fermion structure to backgrounds with different parity and time-reversal assignments. Their later convergence into a common Rashba-like infrared form should therefore be interpreted as an effective universality, not as a microscopic identification.

Operationally, the $\mathfrak g_1$ branch may be viewed as the magnetic-type sector of the theory, while the $\mathfrak g_2$ branch naturally emphasizes the electric-type sector. This is useful not only conceptually but also editorially: it highlights that the effective spin--orbit interaction discussed later is not inserted by hand, but emerges as the common low-energy footprint of two distinct relativistic tensor couplings.

\paragraph{Plane-wave propagation and the relativistic spectral problem.}
A second genuinely relativistic aspect of the theory concerns fermion propagation in constant backgrounds. For uniform fields, Eq.~\eqref{eq:Dirac-compact-rel} admits plane-wave solutions of the form
\begin{equation}
\psi(x)=u(p)e^{-ip\cdot x},
\label{eq:plane-wave-rel}
\end{equation}
where the spinor amplitude obeys
\begin{equation}
\left(
\Gamma^\mu_{\mathrm{eff}}p_\mu-m
\right)u(p)=0.
\label{eq:algebraic-plane-wave-rel}
\end{equation}
The corresponding relativistic dispersion relation follows from the secular equation
\begin{equation}
\det\!\left(\Gamma^\mu_{\mathrm{eff}}p_\mu-m\right)=0.
\label{eq:dispersion-general-rel}
\end{equation}

Equation \eqref{eq:dispersion-general-rel} is the central spectral equation of the relativistic theory. Even before any nonrelativistic reduction is performed, it already shows that the background may deform the usual mass shell, split branches according to spinor structure, and generate anisotropic dependence on the relative orientation between momentum and background.

These features become especially transparent in the static subclasses introduced above. In the magnetic-type branch, with $\mathfrak g_1\neq0$, $\mathfrak g_2=0$, and $F_{i0}=0$, the Dirac equation reduces to
\begin{equation}
\left(
i\gamma^\mu\partial_\mu
+\mathfrak g_1 F_{ij}\gamma^i\gamma_5\,i\partial^j
-m
\right)\psi=0.
\label{eq:Dirac-g1-static-rel}
\end{equation}
To exhibit explicitly the branch splitting, let us choose a constant magnetic field along the $z$ axis,
\begin{equation}
\mathbf B = B\,\hat{\mathbf z},
\qquad
F_{12}=B,
\label{eq:Bz-choice-rel}
\end{equation}
and a plane wave propagating along the $x$ direction,
\begin{equation}
p^\mu=(E,p,0,0).
\label{eq:px-choice-rel}
\end{equation}
Under these conditions, Eq.~\eqref{eq:algebraic-plane-wave-rel} becomes
\begin{equation}
\Big(
\gamma^0 E
-\gamma^1 p
-\eta_B\,p\,\gamma^2\gamma_5
-m
\Big)u(p)=0,
\qquad
\eta_B\equiv \mathfrak g_1 B.
\label{eq:magnetic-example-operator}
\end{equation}
The determinant of the $4\times4$ operator factorizes as
\begin{align}
\det \mathcal D_B
&=
\det\!\Big(
\gamma^0 E
-\gamma^1 p
-\eta_B p\,\gamma^2\gamma_5
-m
\Big)
\nonumber\\
&=
\Big[m^2+p^2(1+\eta_B^2)-E^2-2m\eta_B p\Big]\notag \\ &\times
\Big[m^2+p^2(1+\eta_B^2)-E^2+2m\eta_B p\Big].
\label{eq:det-magnetic-example}
\end{align}
Therefore, the relativistic dispersion relation splits into two branches,
\begin{equation}
E_{\pm}^2
=
m^2+p^2(1+\eta_B^2)\pm 2m\eta_B p.
\label{eq:disp-magnetic-example}
\end{equation}

This explicit example is useful for several reasons. First, it shows concretely that the background deforms the relativistic mass shell. Second, it exhibits branch splitting already at the Dirac level, before any Foldy--Wouthuysen expansion is performed. Third, it makes the directional character of the interaction transparent: the effect is maximal when the momentum is transverse to the magnetic field, whereas for momentum parallel to $\mathbf B$ the tensor contraction vanishes and the spectrum reverts to the undeformed Dirac form. In this sense, the spectral anisotropy is not a secondary detail, but one of the defining relativistic signatures of the model.

\begin{figure}[tbh]
\centering
\includegraphics[width=\columnwidth]{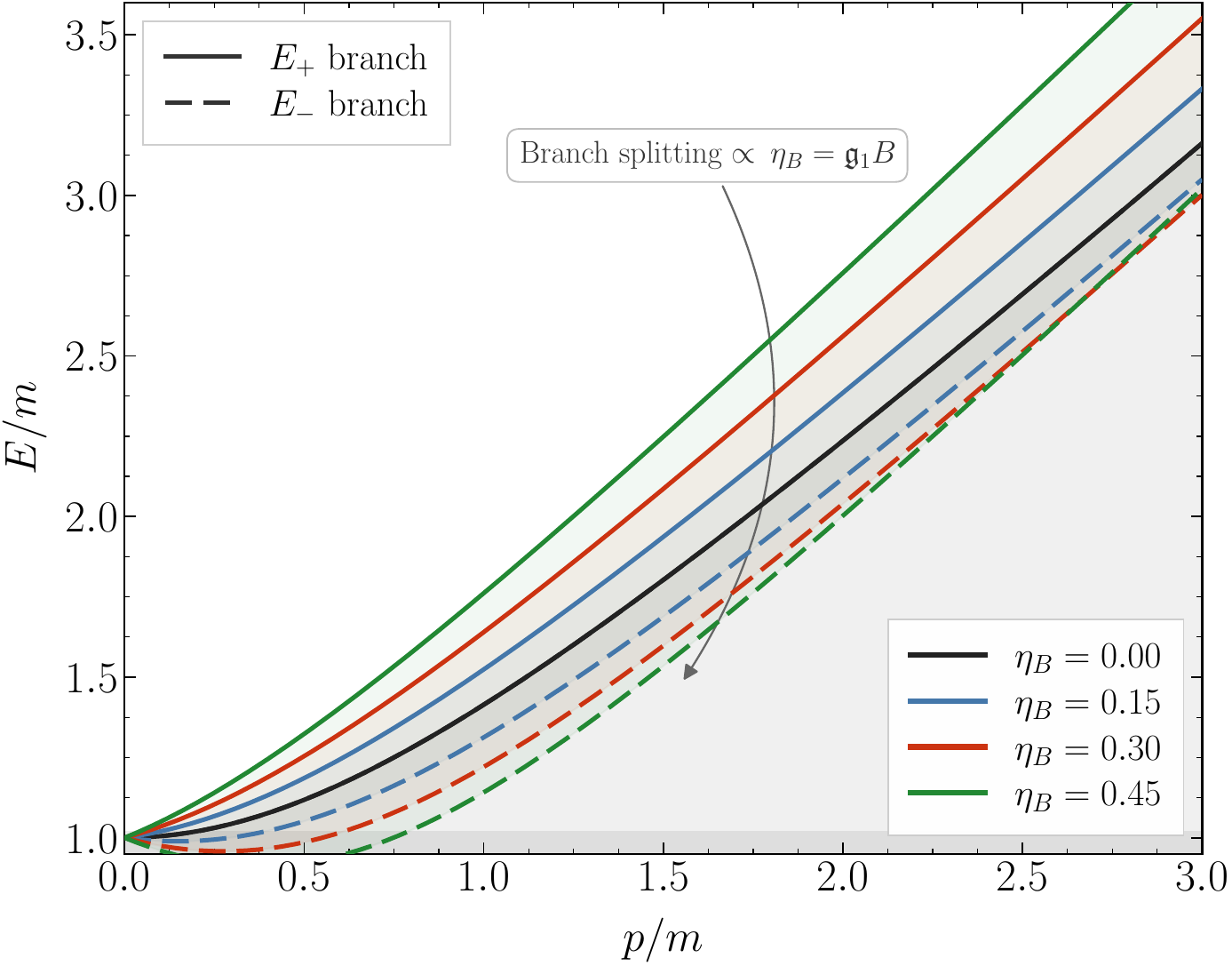}
\caption{Relativistic dispersion relation $E_{\pm}/m$ as a function of the
dimensionless momentum $p/m$ for the magnetic-type branch of the nonminimal
model, Eq.~\eqref{eq:disp-magnetic-example}. Solid and dashed curves
correspond to the $E_{+}$ and $E_{-}$ branches, respectively, for four
representative values of the deformation parameter
$\eta_{B}\equiv\mathfrak{g}_{1}B \in \{0.00,\,0.15,\,0.30,\,0.45\}$.
The shaded gray curve shows the undeformed Dirac dispersion
$E/m=\sqrt{1+(p/m)^{2}}$, recovered in the limit $\eta_{B}\to 0$.
Translucent bands highlight the energy gap between the two branches for each
value of $\eta_{B}$.}
\label{fig:dispersion-magnetic}
\end{figure}
Figure~\ref{fig:dispersion-magnetic} illustrates the two central
predictions of Eq.~\eqref{eq:disp-magnetic-example} at the relativistic level.
First, for any $\eta_{B}\neq 0$ the mass shell splits into two distinct
branches: the $E_{+}$ branch is systematically enhanced with respect to the
free-particle value, while the $E_{-}$ branch is suppressed.
The splitting grows as $2m\eta_{B}p$ at leading order and therefore becomes
increasingly pronounced at large momenta $p\gg m$, where the
deformation term dominates over the mass contribution.
Second, both branches coincide with the standard Dirac curve (gray) at
$p=0$, confirming that the splitting is a purely kinematic effect driven by
the momentum--background tensor contraction: for vanishing spatial momentum
the nonminimal sector contributes nothing to the energy.
Together, these features characterize the branch splitting as a genuine
relativistic signature of the model, already present at the Dirac level
before any Foldy--Wouthuysen reduction is performed. In the nonrelativistic
limit, this splitting descends into the spin-dependent momentum shift that
later governs the effective Rashba coupling on the ring.

The electric-type branch behaves analogously. With $\mathfrak g_1=0$, $\mathfrak g_2\neq0$, $\tilde F_{i0}=0$, and $\mathbf E=E\,\hat{\mathbf z}$, the reduced Dirac operator becomes
\begin{equation}
\left(
i\gamma^\mu\partial_\mu
+\mathfrak g_2 \tilde F_{ij}\gamma^i\gamma_5\,i\partial^j
-m
\right)\psi=0,
\label{eq:Dirac-g2-static-rel}
\end{equation}
and the same algebra yields an analogous branch splitting after the replacement
\begin{equation}
\eta_B=\mathfrak g_1 B
\quad\longrightarrow\quad
\eta_E=\mathfrak g_2 E.
\label{eq:eta-replacement-rel}
\end{equation}
Thus the two sectors are distinct in their microscopic origin but parallel in the way they reorganize relativistic propagation.

\paragraph{Background classification and low-energy bridge.}
From the relativistic perspective, the admissible backgrounds may be organized in terms of the Lorentz invariants
\begin{equation}
\mathcal I_1 = F_{\mu\nu}F^{\mu\nu},
\qquad
\mathcal I_2 = F_{\mu\nu}\tilde F^{\mu\nu},
\label{eq:invariants-rel}
\end{equation}
together with the distinction between electric-type, magnetic-type, and mixed sectors. This classification is useful because it separates what is universal in the relativistic theory from what belongs only to a later lower-dimensional realization. At this stage there is nothing intrinsically ring-like in the theory. The ring will enter only later as an analytically tractable geometry in which the background-induced spin structure becomes especially transparent.

The essential relativistic lesson of the present section is therefore the following. The modified Dirac equation is already a nontrivial background-dependent propagation problem. Its temporal sector selects the admissible one-particle branches, its bilinear structure defines the appropriate effective current, and its plane-wave solutions reveal a deformed relativistic spectrum with directional spin dependence. The low-energy Hamiltonian to be derived in the next section should be read as a controlled descendant of this relativistic kinetic deformation.

In that regime, the background-dependent terms inherited from $\Gamma^\mu_{\mathrm{eff}}$ generate effective operators of the schematic form
\begin{equation}
H_{\mathrm{SO}}\sim
\boldsymbol{\mathcal F}\cdot(\boldsymbol p\times\boldsymbol\sigma),
\label{eq:SO-preview-rel}
\end{equation}
with $\boldsymbol{\mathcal F}$ standing for the effective tensor background associated with either the $F_{\mu\nu}$ or the $\tilde F_{\mu\nu}$ branch. The explicit Hamiltonian construction, the even--odd decomposition, and the Foldy--Wouthuysen reduction leading to Eq.~\eqref{eq:SO-preview-rel} are developed in Sec.~\ref{sec:NR_and_ring}.

\section{Nonrelativistic limit}
\label{sec:NR_and_ring}

In this section we derive the full nonrelativistic limit of the
modified Dirac equation introduced in Sec.~\ref{sec:model}, and then
perform a systematic dimensional reduction to obtain the Hamiltonian
governing the dynamics of a fermion constrained to a one--dimensional
ring of radius $r_{0}$ embedded in the plane.
No approximation beyond the first order in $1/m$ is invoked, and all
terms linear and quadratic in the nonminimal couplings
$\mathfrak{g}_{1}$ and $\mathfrak{g}_{2}$ are retained.

\subsection{Modified Dirac equation and Hamiltonian form}

Our first goal is to cast the relativistic equation into a genuine Hamiltonian form. This step is essential because the Foldy--Wouthuysen procedure distinguishes between operators that act within a fixed energy sector and operators that mix positive- and negative-energy components, and only in Hamiltonian language can that distinction be read off transparently.

A crucial point in the present model is that the derivative couplings
modify not only the spatial part of the Dirac operator, but also its
temporal component. Therefore, before attempting any Hamiltonian or
Foldy--Wouthuysen analysis, one must isolate and invert the operator
multiplying the time derivative, as already stated in Eq~\ref{eq:Heff}. If we now define the quantity
\begin{equation}
W_{i} = -\mathfrak{g}_{1}E_{i} + \mathfrak{g}_{2}B_{i},
\end{equation}
and write
\begin{equation}
\Gamma^0_{\mathrm{eff}} = \beta + K,
\qquad
K = W_{i}\gamma^{i}\gamma_{5},
\label{eq:Gamma0-K-final}
\end{equation}
then the algebraic properties of $K$ are straightforward. One finds
\begin{equation}
[\beta,K]=0,
\end{equation}
and
\begin{equation}
K^{2} = W^{2}\,\mathds{1},
\qquad
W^{2}
=
\mathfrak{g}_{1}^{2}\mathbf{E}^{2}
+\mathfrak{g}_{2}^{2}\mathbf{B}^{2}
-2\mathfrak{g}_{1}\mathfrak{g}_{2}\mathbf{E}\cdot\mathbf{B}.
\label{eq:W2-final}
\end{equation}
Therefore, $\Gamma^0_{\mathrm{eff}}$ can be inverted exactly:
\begin{equation}
(\Gamma^0_{\mathrm{eff}})^{-1}
=
\frac{\beta - K}{1 - W^{2}},
\label{eq:Gamma0-inverse-final}
\end{equation}
provided $1 - W^{2} \neq 0$. This expression shows explicitly that the temporal sector contributes
nontrivially through $(\Gamma^0_{\mathrm{eff}})^{-1}$, and therefore cannot be
neglected in general. The resulting Hamiltonian can be cast in the form
\begin{align}
H_{\mathrm{eff}}
&=
eA_0
+
(\Gamma^0_{\mathrm{eff}})^{-1}\Gamma^j_{\mathrm{eff}}\pi_j
+
(\Gamma^0_{\mathrm{eff}})^{-1}m ,
\\[1mm]
\Gamma^\nu_{\mathrm{eff}}
&=
\gamma^\nu+\Delta^\nu,
\qquad
\Delta^\nu
=
\mathfrak g_1 F_\mu{}^\nu \gamma^\mu\gamma_5
+
\mathfrak g_2 \tilde F_\mu{}^\nu \gamma^\mu\gamma_5 .
\end{align}
This form makes explicit that the temporal sector enters not only
through the electrostatic term $eA_0$, but also through the inverse
operator $(\Gamma^0_{\mathrm{eff}})^{-1}$.

In the regime $|\mathfrak{g}_{i}F_{\mu\nu}|\ll 1$, one may expand
\begin{equation}
(\Gamma^0_{\mathrm{eff}})^{-1}
=
\beta - \Delta^0 + \mathcal O(\mathfrak g^2),
\end{equation}
\begin{align}
H_{\mathrm{eff}}
&=\beta m+\boldsymbol{\alpha}\cdot\boldsymbol{\pi}+eA_0 \notag \\
&+\beta\Delta^j\pi_j-\Delta^0\gamma^j\pi_j-m\Delta^0+\mathcal O(\mathfrak g^2).
\label{eq:Heff-expanded-firstorder}
\end{align}
Equation~\eqref{eq:Heff-expanded-firstorder} is the appropriate
starting point for the Foldy--Wouthuysen analysis.

In physically relevant configurations such as static backgrounds with
$F_{i0}=0$ or $\tilde{F}_{i0}=0$, one has $\Delta^0=0$, implying
\begin{equation}
\Gamma^{0} = \beta,
\end{equation}
and the standard Hamiltonian structure is consistently recovered. This observation justifies the use of conventional nonrelativistic
methods in those sectors, while the general formulation above
provides the complete relativistic framework.

This decomposition already reveals the logic of the problem. The standard Dirac kinetic term $\boldsymbol{\alpha}\cdot\boldsymbol{\pi}$ is odd in the Foldy--Wouthuysen sense and therefore mixes large and small components, whereas the nonminimal pieces may contribute to both diagonal and off-diagonal sectors depending on the background configuration. As a consequence, even weak values of $\mathfrak g_{1}$ and $\mathfrak g_{2}$ can leave nontrivial low-energy traces after the relativistic negative-energy sector is perturbatively removed.

\subsection{Even--odd decomposition}

The even--odd split organizes the Hamiltonian according to its physical effect on the Dirac bispinor. Even operators act within a fixed energy sector and survive directly in the nonrelativistic limit, whereas odd operators mix positive- and negative-energy components and generate effective Pauli and spin--orbit terms only after the Foldy--Wouthuysen transformation is carried out.

Following the standard Foldy--Wouthuysen (FW) procedure, we split
$H$ into even and odd operators with respect to $\beta$. We now decompose the Hamiltonian in the standard form
\begin{equation}
H_{\mathrm{eff}}
=
\beta m+\mathcal E+\mathcal O,
\end{equation}
where
\begin{equation}
[\beta,\mathcal E]=0,
\qquad
\{\beta,\mathcal O\}=0.
\end{equation}

The ordinary Dirac term
\begin{equation}
\boldsymbol{\alpha}\cdot\boldsymbol{\pi}
\end{equation}
is odd, while the correction $-m\Delta^0$ is even, since
$\Delta^0\sim \gamma^i\gamma_5$ commutes with $\beta$.

The term $\beta\Delta^j\pi_j$ contains both even and odd pieces,
depending on the explicit Dirac structure appearing in $\Delta^j$.
Accordingly, we define
\begin{equation}
\beta\Delta^j\pi_j
=
\big(\beta\Delta^j\pi_j\big)_{\rm even}
+
\big(\beta\Delta^j\pi_j\big)_{\rm odd}.
\end{equation}
Since $\Delta^0$ is even and $\gamma^j$ is odd with respect to
$\beta$, the term $\Delta^0\gamma^j\pi_j$ is odd. Therefore, the
even and odd sectors are
\begin{align}
\mathcal E
&=
eA_0
-
m\Delta^0
+
\big(\beta\Delta^j\pi_j\big)_{\rm even},
\label{eq:E-sector}
\\[1mm]
\mathcal O
&=
\boldsymbol{\alpha}\cdot\boldsymbol{\pi}
-
\Delta^0\gamma^j\pi_j
+
\big(\beta\Delta^j\pi_j\big)_{\rm odd}.
\label{eq:O-sector}
\end{align}

These operators already encode the qualitative roles of the two couplings. The scalar products with $\mathbf D$ distort the kinetic sector along preferred background directions, whereas the cross products are the seeds of the effective spin--orbit interaction because they intertwine orbital motion and spin orientation. In that sense, the decomposition above already anticipates which pieces will become Rashba-like after the nonrelativistic reduction.

\subsection{Foldy--Wouthuysen transformation}

The Foldy--Wouthuysen step converts the relativistic operator content into a transparent low-energy Hamiltonian. Once the odd sector is squared, virtual transitions to the small component are encoded as effective operators acting entirely on the positive-energy sector, which is why Pauli, drift, and Rashba-like terms emerge from the same algebraic structure.

Since the equation has now been cast in genuine Schr\"odinger form,
the standard FW expansion can be applied. Up to order $1/m$, one has
\begin{equation}
H_{\mathrm{FW}}
=
\beta m
+\mathcal E
+\frac{\beta}{2m}\mathcal O^2
-\frac{1}{8m^2}[\mathcal O,[\mathcal O,\mathcal E]]
+\cdots.
\label{eq:FW-master-app}
\end{equation}
If one is interested only in the leading nonrelativistic structure,
it is sufficient to retain
\begin{equation}
H_{\mathrm{FW}}
\simeq
\beta m
+\mathcal E
+\frac{\beta}{2m}\mathcal O^2 .
\label{eq:FW-leading-app}
\end{equation}

To evaluate $\mathcal O^2$ systematically, we write
\begin{equation}
\mathcal O=\mathcal O_0+\delta\mathcal O,
\qquad
\mathcal O_0=\boldsymbol{\alpha}\cdot\boldsymbol{\pi},
\end{equation}
with $\delta\mathcal O=\mathcal O(\mathfrak g)$. Then
\begin{equation}
\mathcal O^2
=
\mathcal O_0^2
+
\{\mathcal O_0,\delta\mathcal O\}
+
\mathcal O(\mathfrak g^2).
\label{eq:O2-split}
\end{equation}
The zeroth-order contribution is the usual one,
\begin{equation}
\mathcal O_0^2
=
(\boldsymbol{\alpha}\cdot\boldsymbol{\pi})^2
=
\boldsymbol{\pi}^2
-
e\,\boldsymbol{\Sigma}\cdot\mathbf B,
\label{eq:O0-square}
\end{equation}
where $\Sigma^k=\mathrm{diag}(\sigma^k,\sigma^k)$.

Thus, projecting onto the positive-energy sector, the leading
nonrelativistic Hamiltonian becomes
\begin{align}
H_{\mathrm{NR}}
&=
m
+
eA_0
+
\frac{\boldsymbol{\pi}^2}{2m}
-
\frac{e}{2m}\,\boldsymbol{\sigma}\cdot\mathbf B
\nonumber\\
&\quad
-
m\,\Delta^0_{(+)}
+
\big(\beta\Delta^j\pi_j\big)^{(+)}_{\rm even}
\nonumber\\
&\quad
+
\frac{1}{2m}
\Big\{
\boldsymbol{\alpha}\cdot\boldsymbol{\pi},
-\Delta^0\gamma^j\pi_j
+
\big(\beta\Delta^j\pi_j\big)_{\rm odd}
\Big\}_{(+)}
+\cdots,
\label{eq:HNR-general-final}
\end{align}
where the symbol $(+) $ denotes projection onto the upper two
components. Equation~\eqref{eq:HNR-general-final} is the general low-energy
Hamiltonian associated with the nonminimal derivative model. Its
explicit form depends on the electromagnetic configuration through
$\Delta^\mu$, and in particular on whether the temporal components
$F_{i0}$ and $\tilde F_{i0}$ vanish or not.

Equation above is the master Pauli Hamiltonian derived from the relativistic theory. It contains, in a unified form, the ordinary orbital motion, the Zeeman response, background-induced drifts, genuine spin--orbit couplings, and quadratic corrections in the nonminimal parameters. Every effective ring Hamiltonian discussed below is obtained by projecting this operator onto the azimuthal degree of freedom, so its physical content should be kept in mind throughout the remainder of the paper.

\subsection{Second-order structure.}
Expanding the inverse temporal operator up to quadratic order in the
nonminimal couplings, we obtain
\begin{equation}
(\Gamma^0_{\mathrm{eff}})^{-1}
=
\beta-\Delta^0+\beta(\Delta^0)^2+\mathcal O(\mathfrak g^3).
\end{equation}
Therefore,
\begin{align}
H_{\mathrm{eff}}
&=
\beta m
+\boldsymbol{\alpha}\cdot\boldsymbol{\pi}
+eA_0
+\beta\Delta^j\pi_j
-\Delta^0\gamma^j\pi_j
-m\Delta^0
\nonumber\\
&\quad
-\Delta^0\Delta^j\pi_j
+\beta(\Delta^0)^2\gamma^j\pi_j
+\beta m(\Delta^0)^2
+\mathcal O(\mathfrak g^3).
\end{align}
Accordingly, the even and odd sectors become
\begin{align}
\mathcal E
&=
eA_0-m\Delta^0+\beta m(\Delta^0)^2 \notag \\
&+\big(\beta\Delta^j\pi_j\big)_{\rm even}-\big(\Delta^0\Delta^j\pi_j\big)_{\rm even},
\\
\mathcal O
&=
\boldsymbol{\alpha}\cdot\boldsymbol{\pi}-\Delta^0\gamma^j\pi_j+\big(\beta\Delta^j\pi_j\big)_{\rm odd} \notag\\
&-\big(\Delta^0\Delta^j\pi_j\big)_{\rm odd}+\beta(\Delta^0)^2\gamma^j\pi_j.
\end{align}
The Foldy--Wouthuysen Hamiltonian up to second order in the couplings then reads
\begin{align}
H_{\rm FW}^{(2)}
&=
\beta m+\mathcal E +\frac{\beta}{2m}\left[\mathcal O_0^2+\{\mathcal O_0,\delta\mathcal O_1\}\right]\notag\\
&+\frac{\beta}{2m}\left[\delta\mathcal O_1^2+\{\mathcal O_0,\delta\mathcal O_2\}\right]+\mathcal O(\mathfrak g^3),
\end{align}
where
\begin{align}
\mathcal O_0 &=\boldsymbol{\alpha}\cdot\boldsymbol{\pi},\\
\delta\mathcal O_1
&=
-\Delta^0\gamma^j\pi_j
+\big(\beta\Delta^j\pi_j\big)_{\rm odd},\\
\delta\mathcal O_2
&=
-\big(\Delta^0\Delta^j\pi_j\big)_{\rm odd}
+\beta(\Delta^0)^2\gamma^j\pi_j.
\end{align}

\section{Effective dynamics on a ring}

We now adapt the Pauli Hamiltonian to the geometry relevant for quantum rings. The objective is to rewrite the background couplings in radial and azimuthal components so that the later confinement to a fixed radius can be performed cleanly and with a direct geometric interpretation of each surviving term.

We now restrict the dynamics to the plane $z=0$, writing
$\mathbf{r}=(x,y)$ and allowing for generic fields
$\mathbf{E}=(E_x,E_y,E_z)$ and $\mathbf{B}=(B_x,B_y,B_z)$.
Introducing polar coordinates,
\begin{equation}
x=r\cos\varphi,\qquad y=r\sin\varphi,
\end{equation}
and defining
\begin{align}
E_r &= E_x \cos\varphi + E_y \sin\varphi,\\
E_{\varphi} &= -E_x\sin\varphi + E_y\cos\varphi, \\
B_r &= B_x \cos\varphi + B_y \sin\varphi,\\
B_{\varphi} &= -B_x\sin\varphi + B_y\cos\varphi,
\end{align}
the kinematic operators become
\begin{align}
\pi_x &= \cos\varphi\,\pi_r
      - \frac{\sin\varphi}{r}\,\pi_{\varphi}, \\
\pi_y &= \sin\varphi\,\pi_r
      + \frac{\cos\varphi}{r}\,\pi_{\varphi},
\end{align}
where $\pi_{\varphi}=-i\partial_{\varphi}-eA_{\varphi}$.
The products entering $H_{\rm Pauli}$ then reduce to
\begin{align}
\mathbf{E}\cdot\boldsymbol{\pi}
  &= \frac{E_{\varphi}}{r}\,\pi_{\varphi}, &
\mathbf{B}\cdot\boldsymbol{\pi}
  &= \frac{B_{\varphi}}{r}\,\pi_{\varphi}, \\
(\mathbf{B}\times\boldsymbol{\pi})_{z}
  &= \frac{B_{r}}{r}\,\pi_{\varphi}, &
(\mathbf{E}\times\boldsymbol{\pi})_{z}
  &= \frac{E_{r}}{r}\,\pi_{\varphi}.
\end{align}

At this stage the physical roles of the background components become very transparent. Tangential fields couple directly to the surviving orbital momentum $\pi_{\varphi}$ and therefore behave as effective drifts along the ring, whereas radial components feed the cross products and control the spin-dependent azimuthal transport. This distinction becomes decisive once the radial motion is frozen.

\subsection{Effective Hamiltonian on a ring}

Confinement to a ring converts the reduced Pauli problem into a genuinely mesoscopic system with a single angular degree of freedom. In this geometry the nonminimal background acts as an effective spin-dependent gauge structure, so spectral shifts, phase accumulation, and persistent currents all become controlled by the same azimuthal operator.

We now constrain the fermion to move on a ring of fixed radius $r_{0}$,
implementing $\pi_{r}=0$ and $r\to r_{0}$.  The kinetic term becomes
\begin{equation}
\frac{\pi^{2}}{2m}
=
\frac{1}{2m r_{0}^{2}}\,\pi_{\varphi}^{2}.
\end{equation}
The spin operator is decomposed as
\begin{align}
\sigma_{r} &= \sigma_x\cos\varphi + \sigma_y\sin\varphi,\\
\sigma_{\varphi} &= -\sigma_x\sin\varphi + \sigma_y\cos\varphi.
\end{align}

Collecting all contributions, we obtain the full one-dimensional
Hamiltonian on the ring:
\begin{align}
H_{\rm ring}
=&\;
m + eA_{0}
+\frac{\pi_{\varphi}^{2}}{2m r_{0}^{2}}
-\frac{e}{2m}\left(
   B_{r}\sigma_{r}
 + B_{\varphi}\sigma_{\varphi}
 + B_{z}\sigma_{z}
\right)
\nonumber\\[1mm]
&
-\frac{\mathfrak{g}_{1}}{m r_{0}}\,E_{\varphi}\,\pi_{\varphi}
-\frac{\mathfrak{g}_{2}}{m r_{0}}\,B_{\varphi}\,\pi_{\varphi}
\nonumber\\
&
+\frac{\mathfrak{g}_{1}}{m r_{0}}\,B_{r}\,\pi_{\varphi}\,\sigma_{z}
-\frac{\mathfrak{g}_{2}}{m r_{0}}\,E_{r}\,\pi_{\varphi}\,\sigma_{z}
\nonumber\\
&
+\frac{1}{2m r_{0}^{2}}
\Big[
\mathfrak{g}_{1}^{2}(E_{\varphi}^{2} + r_{0}^{2} m^{2} B_{r}^{2})
+\mathfrak{g}_{2}^{2}(B_{\varphi}^{2} + r_{0}^{2} m^{2} E_{r}^{2})
\nonumber\\
&
+2\mathfrak{g}_{1}\mathfrak{g}_{2}
 (E_{\varphi}B_{\varphi} + r_{0}^{2}m^{2}E_{r}B_{r})
\Big]D_{\varphi}^{2}.
\label{eq:H_ring_final}
\end{align}

This full ring Hamiltonian shows explicitly how the external backgrounds survive the confinement. Terms proportional to $\pi_{\varphi}$ without Pauli matrices produce spin-blind drifts, while the coefficients multiplying $\pi_{\varphi}\sigma_z$ split the two spin sectors and are therefore the precursors of the Rashba-like coupling analyzed later. The quadratic derivative term acts as a background-induced renormalization of the effective inertia of the azimuthal motion.

It is convenient to define the dimensionless coefficients
\begin{align}
\xi_{1} &= m r_{0}\mathfrak{g}_{1} B_{r} - \mathfrak{g}_{2}E_{r}, \\
\xi_{2} &= - m r_{0}(\mathfrak{g}_{1} E_{\varphi} + \mathfrak{g}_{2}B_{\varphi}),
\end{align}
so that the Hamiltonian can be written in the compact form
\begin{align}
H_{\rm ring}
&=
\frac{\pi_{\varphi}^{2}}{2m r_{0}^{2}}
+ \frac{\xi_{1}}{m r_{0}}\,\pi_{\varphi}\sigma_{z}
+ \frac{\xi_{2}}{m r_{0}}\,\pi_{\varphi}\notag\\
&+ H_{\rm quad}
+ H_{\rm Zeeman}
+ m + eA_{0},
\end{align}
with
\begin{align}
H_{\rm quad}&=\frac{(\mathfrak{g}_{1} E_{\varphi}+\mathfrak{g}_{2}B_{\varphi})^{2}}{2m r_{0}^{2}}D_{\varphi}^{2}+\notag\\
            &+\frac{(m r_{0} \mathfrak{g}_{1} B_{r} - \mathfrak{g}_{2}E_{r})^{2}}{2m r_{0}^{2}}D_{\varphi}^{2},\\
H_{\rm Zeeman}
&=
-\frac{e}{2m}
\left(
   B_{r}\sigma_{r}
 + B_{\varphi}\sigma_{\varphi}
 + B_{z}\sigma_{z}
\right).
\end{align}

The parametrization in terms of $\xi_{1}$ and $\xi_{2}$ is physically useful because it separates the genuinely spin-sensitive momentum shift from the spin-independent orbital drift. In the one-dimensional problem, the most distinctive mesoscopic signatures arise from $\xi_{1}$, since only this combination displaces opposite spin branches relative to one another.

In the special case of a static planar magnetic configuration with
$E_{r}=E_{\varphi}=0$ and $B_{r}=\mathcal{F}_{12}/e$, the Hamiltonian
\eqref{eq:H_ring_final} reduces to
\begin{equation}
H_{\rm ring}
=
\frac{1}{2m r_{0}^{2}}
\left(
 i\partial_{\varphi}
 + \xi\sigma_{r}
\right)^{2}
- \xi^{2},
\end{equation}
which is precisely the Rashba-like hermitian Hamiltonian analyzed in Sec.~\ref{sec:quantum-rings}.

This compact form is especially illuminating: the background enters exactly as a spin-dependent gauge potential inside the covariant angular momentum. The parameter $\xi$ therefore controls, in one stroke, the spectral displacement of the spin branches, the geometric phase accumulated over one revolution, and the persistent spin current supported by the ring. The subtractive constant $-\xi^{2}$ only fixes the energy zero and does not affect the eigenstates.

\subsection{Model 1}

The general structure introduced in the previous section contains several derivative, chirality--dependent couplings that considerably complicate the dynamics. In order to extract concrete physical consequences, we now focus on a specific subsector in which only the $\mathfrak{g}_{1}$ coupling is present. The Dirac equation then reduces to
\begin{equation}
\left(
  i\gamma^{\mu}\partial_{\mu}
  + \mathfrak{g}_{1} F_{\mu\nu}\gamma^{\mu}\gamma_{5} i\partial^{\nu}
  - m
\right)\psi = 0 .
\end{equation}

To isolate the relevant terms, we split temporal and spatial components and reorganize the equation in a convenient form: The temporal components must be isolated because they determine whether the modified Dirac operator preserves the standard first-order evolution problem. If they remained active, the background would multiply the time derivative itself and the construction of a Hermitian single-particle Hamiltonian would become considerably more subtle.

\begin{align}
\left( \gamma^{0} p^{0} + \mathfrak{g}_{1} F_{i0}\gamma^{i}\gamma_{5} p^{0} \right)\psi
&= \gamma^{i}p^{i}\psi- \mathfrak{g}_{1}F_{0i}\gamma^{0}\gamma_{5}p^{i}\psi \notag\\
   &- \mathfrak{g}_{1} F_{ij}\gamma^{i}\gamma_{5}p^{j}\psi+ m\psi .
\end{align}

To avoid a non-Hermitian Hamiltonian---and consequently a nonunitary time evolution---we assume
\begin{equation*}
F_{i0} = -F_{0i} = 0 ,
\end{equation*}
a condition often used in the literature.\footnote{Alternatively, one could perform a field redefinition $\psi = B\Psi$ such that the dependence on $\partial_{0}\Psi$ mimics the usual Dirac form. For the purposes of the present discussion, simply setting $F_{i0}=0$ suffices.}

Physically, this restriction amounts to focusing on stationary backgrounds for which the nonminimal sector does not modify the coefficient of the first time derivative in the Dirac equation. This is the simplest subsector in which the one-particle Hamiltonian remains manifestly Hermitian without the introduction of additional auxiliary fields.

Under this assumption, the Hamiltonian eigenvalue equation becomes
\begin{equation}
E
\begin{pmatrix}
\varphi \\[2pt]
\chi
\end{pmatrix}
=
\gamma^{0}
\left(
  \gamma^{i}p^{i}
  - \mathfrak{g}_{1} F_{ij}\gamma^{i}\gamma_{5}p^{j}
  + m
\right)
\begin{pmatrix}
\varphi \\[2pt]
\chi
\end{pmatrix}.
\end{equation}

If the new interaction is small compared to $|m\chi|$, the nonrelativistic limit for the large component $\varphi$ leads to the effective Hamiltonian
\begin{equation}
H = \frac{p^{2}}{2m} + \mathfrak{g}_{1} F_{ij} p^{i} \sigma^{j}.
\end{equation}
Using the identity $F_{ij} = \varepsilon_{ijk} B^{k}$, this may be rewritten as
\begin{equation}
H = \frac{p^{2}}{2m}
  + \mathfrak{g}_{1}\,\boldsymbol{B}\cdot (\boldsymbol{p}\times\boldsymbol{\sigma}).
\end{equation}

The final structure is physically remarkable because the magnetic field no longer acts only through a Zeeman alignment term. Instead, it mediates a coupling between orbital motion and spin texture, so a moving fermion experiences Rashba-like precession even without the usual structural-inversion mechanism of condensed-matter systems. This is one of the genuinely new conceptual points of the present model.

This result is noteworthy: even in the absence of electric fields, the derivative coupling induced by $F_{\mu\nu}$ generates a spin--orbit--like structure reminiscent of the Rashba interaction. In what follows, this effective Hamiltonian will serve as one of the two fundamental ingredients for the analogue Rashba dynamics on quantum rings.

\subsection{Model 2}

We now turn to the second sector contained in the general Lagrangian, namely the contribution controlled by the coupling $\mathfrak{g}_{2}$ and involving the dual field tensor $\tilde{F}_{\mu\nu}$. The corresponding Dirac equation reads
\begin{equation}
\left(
  i\gamma^{\mu}\partial_{\mu}
  + \mathfrak{g}_{2}\tilde{F}_{\mu\nu}\gamma^{\mu}\gamma_{5} i\partial^{\nu}
  - m
\right)\psi = 0 .
\end{equation}

Proceeding as in the previous subsection, we separate temporal and spatial parts to obtain
Because the dual tensor interchanges electric and magnetic roles, the same algebra now singles out the static electric sector as the source of the effective spin--orbit coupling. This is why the two models, although microscopically different, become equivalent at the level of the one-dimensional mesoscopic observables.

\begin{align}
\left( \gamma^{0}p^{0} + \mathfrak{g}_{2}\tilde{F}_{i0}\gamma^{i}\gamma_{5} p^{0} \right)\psi
&= \gamma^{i}p^{i}\psi
   - \mathfrak{g}_{2}\tilde{F}_{0i}\gamma^{0}\gamma_{5}p^{i}\psi \notag \\
&\quad -\mathfrak{g}_{2}\tilde{F}_{ij}\gamma^{i}\gamma_{5}p^{j}\psi
   + m\psi .
\end{align}

As before, in order to preserve Hermiticity and avoid nonunitary contributions from first time derivatives, we impose
\begin{equation*}
\tilde{F}_{i0} = -\tilde{F}_{0i} = 0 .
\end{equation*}
This choice isolates the static electric subsector of the dual interaction and guarantees that the effective single-particle problem obtained below admits a Hermitian Schr\"odinger operator.
Under this condition, the Hamiltonian eigenvalue equation becomes
\begin{equation}
E\psi =
\gamma^{0}
\left(
  \gamma^{i}p^{i}
  - \mathfrak{g}_{2}\tilde{F}_{ij}\gamma^{i}\gamma_{5}p^{j}
  + m
\right)\psi ,
\end{equation}
from which the nonrelativistic limit for the large component follows analogously to the previous case:
\begin{equation}
H = \frac{p^{2}}{2m} + \mathfrak{g}_{2}\tilde{F}_{ij}p^{i}\sigma^{j}.
\end{equation}

Using the standard identity for the dual tensor in $(3+1)$ dimensions,
\begin{equation*}
\tilde{F}_{ij} = -\varepsilon_{ijk} E^{k},
\end{equation*}
the Hamiltonian reduces to
\begin{equation}
H = \frac{p^{2}}{2m}
    - \mathfrak{g}_{2}\,\boldsymbol{E}\cdot (\boldsymbol{p}\times\boldsymbol{\sigma}).
\end{equation}

Hence the dual interaction reproduces the same low-energy operator structure as Model~1, but now the electric background drives the spin precession. The observables studied below therefore depend primarily on the effective tensor strength that shifts the azimuthal momentum, not on whether that tensor originated from $F_{\mu\nu}$ or from its dual.

This result shows that the dual sector generates an effective spin--orbit--type coupling driven by the electric field rather than the magnetic field. Both models therefore share the same structural form but with different physical origins. In what follows, we shall unify these two cases into a single effective description suitable for the analysis of quantum rings.

\section{One-dimensional quantum rings}
\label{sec:quantum-rings}

Having obtained the effective nonrelativistic sectors, we now turn to the mesoscopic problem where their geometric content becomes most transparent. A quantum ring is the natural arena for this purpose because the closed trajectory forces the spinor to accumulate phase over a complete cycle, making spectral shifts, geometric phases, and persistent currents direct manifestations of the same effective gauge structure.

\subsection{The analogue Rashba interaction}

The two nonrelativistic Hamiltonians obtained in the previous subsections share an identical operator structure. It is therefore convenient to unify them into a single effective description. We write
\begin{equation*}
H = \frac{p^{2}}{2m} + \mathcal{F}_{ij} p^{i}\sigma^{j},
\end{equation*}
where the tensor $\mathcal{F}_{ij}$ takes the value $\mathfrak{g}_{1}F_{ij}$ for Model~1 and $\mathfrak{g}_{2}\tilde{F}_{ij}$ for Model~2. The interaction term can be decomposed as

This compact notation isolates the universal part of the dynamics. Once the problem is written in terms of $\mathcal F_{ij}$, the later derivations apply simultaneously to the magnetic realization of Model~1 and the electric realization of Model~2, because both are encoded in the same antisymmetric coupling between momentum and spin.

\begin{align}
\mathcal{F}_{ij} p^{i}\sigma^{j}
&= \mathcal{F}_{12}\!\left( p^{1}\sigma^{2} - p^{2}\sigma^{1} \right)
  + \mathcal{F}_{13}\!\left( p^{1}\sigma^{3} - p^{3}\sigma^{1} \right) \notag\\
&\quad + \mathcal{F}_{23}\!\left( p^{2}\sigma^{3} - p^{3}\sigma^{2} \right).
\end{align}

We now restrict the dynamics to the two-dimensional plane and impose
\begin{equation*}
p^{3} = 0, \qquad \mathcal{F}_{23} = \mathcal{F}_{13} = 0 ,
\end{equation*}
consistent with the motion of an electron confined to a ring of fixed radius $r = r_{0}$. In polar coordinates, the linear momentum components read
\begin{equation}
p^{1} = \frac{\sin\varphi}{r_{0}} \left( i\,\frac{\partial}{\partial\varphi} \right),
\qquad
p^{2} = -\frac{\cos\varphi}{r_{0}} \left( i\,\frac{\partial}{\partial\varphi} \right),
\end{equation}
so that the spin--momentum term becomes
\begin{equation}
\mathcal{F}_{ij} p^{i}\sigma^{j}
 = \frac{\mathcal{F}_{12}}{r_{0}}
   \left( \sigma^{2}\sin\varphi + \sigma^{1}\cos\varphi \right)
   \left( i\,\frac{\partial}{\partial\varphi} \right).
\end{equation}

Only the antisymmetric component $\mathcal F_{12}$ survives the planar reduction, which means that the one-dimensional dynamics is controlled by the effective tensor component normal to the ring plane. This is directly analogous to the standard Rashba problem, where an out-of-plane structure induces in-plane spin precession.

Thus, the Hamiltonian in polar coordinates takes the form
\begin{equation*}
H = \frac{1}{2mr_{0}^{2}}
      \left( i\,\frac{\partial}{\partial\varphi} \right)^{2}
    + \frac{\mathcal{F}_{12}}{r_{0}}
      \left( \sigma^{2}\sin\varphi + \sigma^{1}\cos\varphi \right)
      \left( i\,\frac{\partial}{\partial\varphi} \right).
\end{equation*}

The second term clearly reproduces a Rashba-type spin--orbit interaction. An important physical observation emerges here: in our framework, the analogue Rashba coupling can originate not only from an electric field (as in the usual condensed-matter scenario) but also from a magnetic field through the antisymmetric tensor structure.

From this point onward, the spectral, geometric, and thermodynamic analysis will be expressed in terms of the effective tensor $\mathcal{F}_{ij}$, so that the final results apply equally to the magnetic realization of Model~1 and to the electric realization of Model~2.

To obtain a Hermitian Hamiltonian, we follow the procedure of Ref.~\cite{CASANA2015171}, which yields
\begin{equation}
H = \frac{1}{2mr_{0}^{2}}
      \left( i\,\frac{\partial}{\partial\varphi} \right)^{2}
  + \frac{\mathcal{F}_{12}}{r_{0}}
      \sigma_{\rho}\left( i\,\frac{\partial}{\partial\varphi} \right)
  + \frac{i}{2}\frac{\mathcal{F}_{12}}{r_{0}}\,\sigma_{\varphi},
\label{eq:HH}
\end{equation}
where the azimuthal Pauli matrices are defined as

The term proportional to $\sigma_{\varphi}$ is not an ad hoc correction. It is the geometric contribution required by Hermiticity in curvilinear coordinates, and it encodes the fact that the local spin basis rotates as the particle moves around the ring. In modern language, it plays the role of a spin connection associated with the moving frame.

\begin{eqnarray*}
\sigma_{\rho}
  &=& \sigma^{1}\sin\varphi - \sigma^{2}\cos\varphi, \\
\sigma_{\varphi}
  &=& \sigma^{1}\cos\varphi + \sigma^{2}\sin\varphi.
\end{eqnarray*}

These matrices satisfy the identities
\begin{equation*}
\tilde{\sigma}_{\rho} = -\partial_{\varphi}\tilde{\sigma}_{\varphi},
\qquad
\tilde{\sigma}_{\varphi} = \partial_{\varphi}\tilde{\sigma}_{\rho},
\end{equation*}
and admit the explicit matrix representations
\begin{eqnarray*}
\sigma_{\rho}
&=&
\begin{pmatrix}
0 & i e^{-i\varphi} \\
- i e^{i\varphi} & 0
\end{pmatrix}, \\
\sigma_{\varphi}
&=&
\begin{pmatrix}
0 & e^{-i\varphi} \\
e^{i\varphi} & 0
\end{pmatrix}.
\end{eqnarray*}

Before solving the eigenvalue problem, it is convenient to rewrite the Hamiltonian in an alternative but equivalent form, as discussed in Ref.~\cite{CASANA2015171}. One observes that
\begin{equation}
\left( i\frac{\partial}{\partial\varphi} + \xi\sigma_{\rho} \right)^{2}
= \left( i\frac{\partial}{\partial\varphi} \right)^{2}
  + 2\xi\sigma_{\rho}\left( i\frac{\partial}{\partial\varphi} \right)
  + i\xi\sigma_{\varphi} + \xi^{2},
\end{equation}
where $\xi = m r_{0}\mathcal{F}_{12}$. Using this identity, Eq.~\eqref{eq:HH} can be recast as
\begin{equation}
H = \frac{1}{2mr_{0}^{2}}
  \left[
    \left( i\frac{\partial}{\partial\varphi} + \xi\sigma_{\rho} \right)^{2}
    - \xi^{2}
  \right].
\end{equation}

This representation significantly simplifies the determination of the spectrum and will serve as the starting point for the next subsections.

Written in this way, the Hamiltonian is mathematically equivalent to a particle coupled to a non-Abelian, spin-dependent gauge potential. The operator $\xi\sigma_{\rho}$ shifts the azimuthal momentum in opposite ways for different spin branches, which is why the same parameter $\xi$ will control the spectral splitting, the geometric phase, and the persistent spin current.

In order to determine the full energy spectrum and the corresponding spinor eigenstates, we begin by solving the first-order equation generated by the alternative Hamiltonian form:
\begin{equation}
\left( i\,\frac{\partial}{\partial\varphi} + \xi\sigma_{\rho} \right)\psi
 = \varepsilon \psi .
\end{equation}

It is advantageous to solve this first-order problem before returning to the quadratic Hamiltonian. The eigenvalues $\varepsilon$ are the effective angular-momentum shifts induced by the background, and once they are known the physical energy follows immediately from the square of the operator.
In matrix form, using the explicit expression for $\sigma_{\rho}$, this reads
\begin{equation}
\begin{pmatrix}
 i\,\frac{\partial}{\partial\varphi} &
 i\xi e^{-i\varphi} \\
 -i\xi e^{i\varphi} &
 i\,\frac{\partial}{\partial\varphi}
\end{pmatrix}
\begin{pmatrix}
\psi_{\uparrow} \\
\psi_{\downarrow}
\end{pmatrix}
 = \varepsilon
\begin{pmatrix}
\psi_{\uparrow} \\
\psi_{\downarrow}
\end{pmatrix}.
\label{eq:H1}
\end{equation}

Guided by the angular dependence of the coupling, the eigenfunctions are sought in the general form
\begin{equation}
\psi_{n,s}^{\lambda}
 = e^{i\lambda n\varphi}
   \begin{pmatrix}
     A_{\lambda,s}\,e^{-i\varphi/2} \\
     B_{\lambda,s}\,e^{i\varphi/2}
   \end{pmatrix},
\label{eq:Sol1}
\end{equation}
where $n$ is the angular-momentum quantum number, $\lambda=\pm$ indicates the propagation direction, and $s=\pm$ labels the two spin branches. Substituting this ansatz into Eq.~\eqref{eq:H1} yields the eigenvalue condition
\begin{equation}
\varepsilon_{n,s}^{\lambda}
 = \frac{s}{2}\sqrt{1+4\xi^{2}} - \lambda n .
\end{equation}

The square-root factor measures how strongly the spin texture is tilted by the effective coupling. For $\xi=0$ one recovers the uncoupled ring structure, whereas finite $\xi$ continuously displaces the two spin branches and opens the characteristic Rashba-like splitting.

Since the energy eigenvalues of the original Hamiltonian follow from the squared operator, we obtain
\begin{equation}
E_{n,s}^{\lambda}
 = \Omega\left[
      \left(n - \frac{\lambda s}{2}\sqrt{1+4\xi^{2}}\right)^{2}
      - \xi^{2}
   \right],
\end{equation}
with $\Omega = (2mr_{0}^{2})^{-1}$. In the thermodynamic analysis to be presented later, it is convenient to shift
\begin{equation}
n \rightarrow n+\tfrac{1}{2},
\end{equation}
so that $n\in\mathbb{Z}$. Under this redefinition, the spectrum becomes
\begin{equation}
E_{n,s}^{\lambda}
 = \Omega\left[
    \left(
       n + \frac{1}{2}
       - \frac{\lambda s}{2}\sqrt{1+4\xi^{2}}
    \right)^{2}
    - \xi^{2}
   \right].
\end{equation}

This spectrum makes the mesoscopic meaning of the coupling explicit. The parameter $\xi$ does not merely add a constant splitting; it shifts the centers of the angular-momentum parabolas themselves. Consequently, changes in $\xi$ reorganize the ordering of low-lying states and leave clear fingerprints in all transport and thermodynamic quantities.

\bigskip

\noindent\textbf{Eigenfunctions.}
To determine the spinor structure explicitly, we solve the secular equation obtained from Eq.~\eqref{eq:H1}. For concreteness, we evaluate the case $\lambda=+$. Using Eq.~\eqref{eq:Sol1}, the condition relating the spinor components becomes
\begin{equation}
B_{+,s}
 = \frac{1}{2\xi}
   \left( \frac{s}{\cos\theta} - 1 \right)
   A_{+,s},
\end{equation}
where the mixing angle $\theta$ satisfies
\begin{equation}
\cos\theta = \frac{1}{\sqrt{1+4\xi^{2}}}.
\end{equation}

To construct normalized spinors, we choose
\begin{equation}
A_{+,+} = \cos\frac{\theta}{2},
\qquad
B_{+,+} = \sin\frac{\theta}{2},
\end{equation}
which implies the constraint
\begin{equation}
\tan\theta = 2\xi .
\end{equation}
For the $s=-$ branch, we take
\begin{equation}
A_{+,-} = -\sin\frac{\theta}{2}, \qquad
B_{+,-} = \cos\frac{\theta}{2}.
\end{equation}

The resulting properly normalized eigenspinors for $\lambda=+$ are
\begin{align}
\psi_{n,+}^{+}
 &= e^{in\varphi}
   \begin{pmatrix}
     \cos\frac{\theta}{2}\,e^{-i\varphi/2} \\
     i\sin\frac{\theta}{2}\,e^{i\varphi/2}
   \end{pmatrix},
\\
\psi_{n,-}^{+}
 &= e^{in\varphi}
   \begin{pmatrix}
     -\sin\frac{\theta}{2}\,e^{-i\varphi/2} \\
     i\cos\frac{\theta}{2}\,e^{i\varphi/2}
   \end{pmatrix}.
\end{align}

Similarly, for $\lambda=-$ one finds
\begin{align}
\psi_{n,+}^{-}
 &= e^{-in\varphi}
   \begin{pmatrix}
     \cos\frac{\theta}{2}\,e^{-i\varphi/2} \\
     i\sin\frac{\theta}{2}\,e^{i\varphi/2}
   \end{pmatrix},
\\
\psi_{n,-}^{-}
 &= e^{-in\varphi}
   \begin{pmatrix}
     -\sin\frac{\theta}{2}\,e^{-i\varphi/2} \\
     i\cos\frac{\theta}{2}\,e^{i\varphi/2}
   \end{pmatrix}.
\end{align}

These eigenfunctions provide the complete basis for the evaluation of geometric phases, spin currents, and thermodynamic properties developed in the subsequent sections.
They also show that the spin is neither purely radial nor purely tangential. Instead, it is mixed by the angle $\theta$, whose dependence on $\xi$ defines a nontrivial spin texture around the ring. That geometric spin texture is precisely what later feeds the Aharonov--Anandan phase and the angular modulation of the transverse spin currents.

\subsection{Induction of geometric phases}

The phase analysis is the natural continuation of the spectral problem because the effective coupling behaves as a spin-dependent gauge field on a closed path. Once the electron completes one revolution around the ring, the spatial configuration is periodic, but the spinor need not return to itself trivially; the resulting mismatch is precisely what the Aharonov--Anandan phase measures.

We now investigate whether the effective spin--orbit coupling induces nontrivial geometric phases for the eigenstates obtained in the previous subsection. To this end we employ the Aharonov--Anandan (AA) phase, which generalizes the Berry phase to arbitrary cyclic evolutions.

For a normalized eigenstate $\psi_{n,s}^{\lambda}(\varphi)$, the AA geometric phase accumulated over a $2\pi$ rotation around the ring is defined as
\begin{equation}
\Phi_{\mathrm{AA}}^{(\lambda,s)}
 = \int_{0}^{2\pi}
   \psi_{n,s}^{\lambda\dagger}
   \, i\frac{d}{d\varphi}
   \psi_{n,s}^{\lambda}\, d\varphi .
\label{eq:PhiAA}
\end{equation}
The dynamical phase associated with the effective spin--orbit interaction,
\begin{equation}
H_{\mathrm{eff}} = \mathcal{F}_{ij} p^{i}\sigma^{j},
\end{equation}
is given by
\begin{equation}
\Phi_{\mathrm{dyn}}^{(\lambda,s)}
 = -\int_{0}^{2\pi}
    \left[\psi_{n,s}^{\lambda}(\varphi)\right]^{\dagger}
    H_{\mathrm{eff}}
    \psi_{n,s}^{\lambda}(\varphi)\, d\varphi .
\end{equation}
The total phase acquired in one full revolution is therefore
\begin{equation}
\Phi_{\mathrm{Total}}^{(\lambda,s)}
 = \Phi_{\mathrm{AA}}^{(\lambda,s)}
   + \Phi_{\mathrm{dyn}}^{(\lambda,s)}.
\end{equation}

Separating the total phase into geometric and dynamical pieces is physically useful. The dynamical contribution tracks the local interaction energy accumulated along the path, whereas the geometric contribution depends only on how the spinor explores the projective Hilbert space during the cyclic evolution.

Using the explicit eigenfunctions derived earlier, the AA phase evaluates to
\begin{equation}
\Phi_{\mathrm{AA}}^{(\lambda,s)}
 = -2\lambda\pi\left(
      n - \frac{\lambda s}{2}\cos\theta
   \right),
\end{equation}
where the mixing angle $\theta$ satisfies $\cos\theta = (1+4\xi^{2})^{-1/2}$.

Likewise, the dynamical phase contribution is found to be
\begin{equation}
\Phi_{\mathrm{dyn}}^{(+,+)}
 = 2s\pi\xi\sin\theta .
\end{equation}

The two phase contributions depend on the same mixing angle in complementary ways: the geometric term is weighted by $\cos\theta$, while the dynamical part is weighted by $\sin\theta$. This reflects the fact that the effective Rashba field both tilts the spin texture and changes the energy cost of transporting it around the ring.

Combining the two contributions, and recalling that the energy is obtained from the square of the effective angular momentum operator, one may write the energy spectrum compactly as
\begin{equation*}
E_{n,s}^{\lambda}
 = \Omega\left[
      \left(
        \frac{\Phi_{\mathrm{Total}}^{(\lambda,s)}}{2\pi}
      \right)^{2}
      - \xi^{2}
   \right].
\end{equation*}

\begin{figure}[tbh]
\centering
\includegraphics[scale=0.5]{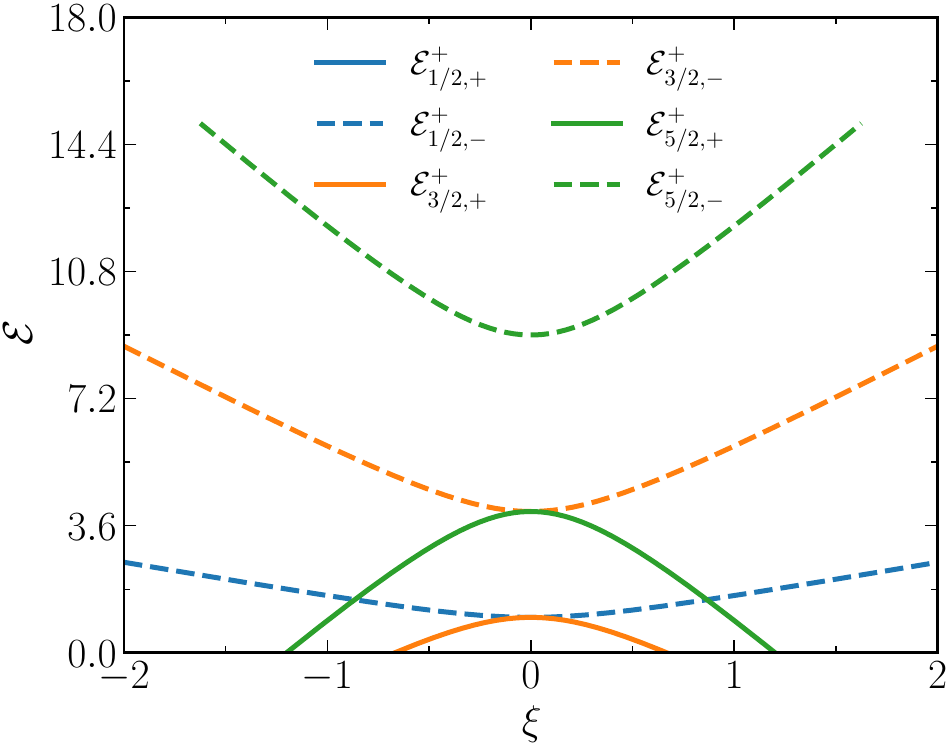}
\caption{Single-particle energy spectrum as a function of the effective Rashba-like coupling $\xi$ for representative branches labeled by $(n,s,\lambda)$. Finite $\xi$ lifts the degeneracies present in the uncoupled ring and displaces the parabolic branches by different amounts, making explicit how the spin-dependent gauge shift reorganizes the low-energy spectrum.}
\label{Fig:Total-Energy}
\end{figure}
Figure~\ref{Fig:Total-Energy} makes the branch rearrangement produced by the coupling very clear. Near $\xi=0$ the levels organize into the familiar ring multiplets, while finite $\xi$ lifts the degeneracies and separates the states according to their spin and propagation labels. The lower branches are driven downward because the effective gauge shift reduces the kinetic cost of selected angular-momentum sectors, whereas the upper branches move upward. This redistribution of levels is the spectral origin of the thermodynamic and transport responses discussed below.

The expression above highlights a key interpretation:
the analogue Rashba coupling introduces a geometric shift in the quantization of angular momentum, encoded in the total geometric phase. This mechanism underlies the modification of the spectrum and governs the behavior of persistent spin currents in the subsequent analysis.

\subsection{Spin currents}

Persistent currents provide the most direct transport manifestation of the effective gauge structure. Even though time-reversal symmetry forbids a net charge current in the present setting, the spin sector can still support stationary transport because opposite spin projections experience opposite momentum shifts around the ring.

In systems with time-reversal symmetry, persistent charge currents vanish identically. Nevertheless, spin--orbit interactions can still support nonvanishing persistent \emph{spin} currents. For the present model, the spin current density along the azimuthal direction is defined as
\begin{equation}
\mathcal{J}_{\varphi}^{a}
 = \frac{1}{2}
   \psi_{n,s}^{\lambda\dagger}
   \left\{ \mathbf{v}_{\varphi},\,\mathbf{s}^{a} \right\}
   \psi_{n,s}^{\lambda},
\end{equation}
with $\mathbf{s}^{a}=\sigma^{a}/2$ and where the velocity operator follows from the commutator
\begin{equation}
\mathbf{v}_{\varphi}
 = i r_{0}[H,\varphi]
 = \frac{1}{i m r_{0}}
     \frac{\partial}{\partial\varphi}
   - \mathcal{F}_{12}\sigma_{\rho}.
\end{equation}

The symmetrized definition is important because velocity and spin fail to commute once spin--orbit coupling is present. The velocity operator itself contains an anomalous term proportional to $\sigma_{\rho}$, which is the direct transport imprint of the effective Rashba field: even states with fixed orbital quantum number acquire a spin-dependent azimuthal drift.

A general expression for the spin current carried by a single eigenstate is
\begin{align*}
\mathcal{J}_{\varphi}^{a}
&= \frac{1}{2mr_{0}}
   \left( \lambda n - \frac{1}{2} \right)
   \left[
      |A_{\lambda,s}|^{2}\sigma^{a}_{11}
      + B_{\lambda,s}^{*} A_{\lambda,s}
        e^{-i\varphi}\sigma^{a}_{21}
   \right]
   \\
&\quad
 + \frac{1}{2mr_{0}}
   \left( \lambda n + \frac{1}{2} \right)
   \left[
      A_{\lambda,s}^{*} B_{\lambda,s}
        e^{i\varphi}\sigma^{a}_{12}
      + |B_{\lambda,s}|^{2}\sigma^{a}_{22}
   \right]
\\
&\quad
 - \frac{\mathcal{F}_{12}}{2}
   \left( \delta_{x,a}\cos\varphi + \delta_{y,a}\sin\varphi \right),
\end{align*}
where $a=\{x,y,z\}$ and the last term originates from the spin--orbit-induced modification of the velocity operator.
This expression separates the current into an orbital contribution, controlled by the winding number $n$, and a genuinely spin--orbit contribution, controlled by $\mathcal F_{12}$. Their competition determines the sign and magnitude of each spin component and explains why the longitudinal and transverse currents behave differently.

\bigskip

\noindent\textbf{Spin current in the $z$ direction.}
The degenerate eigenstates at each energy level allow for nontrivial superpositions. The physically relevant quantity is obtained by summing the contributions from the four degenerate states in the totally symmetric combination. Doing so yields
\begin{align}
&\mathcal{J}_{\varphi}^{z}
= \frac{1}{4}
   \psi_{n,+}^{+\dagger}
     \frac{1}{2}\{\mathbf{v}_{\varphi},\mathbf{s}^{z}\}
     \psi_{n,+}^{+}
 + \frac{1}{4}
   \psi_{n,-}^{-\dagger}
     \frac{1}{2}\{\mathbf{v}_{\varphi},\mathbf{s}^{z}\}
     \psi_{n,-}^{-}
\notag\\
& + \frac{1}{4}
   \psi_{-n,-}^{+\dagger}
     \frac{1}{2}\{\mathbf{v}_{\varphi},\mathbf{s}^{z}\}
     \psi_{-n,-}^{+}
 + \frac{1}{4}
   \psi_{-n,+}^{-\dagger}
     \frac{1}{2}\{\mathbf{v}_{\varphi},\mathbf{s}^{z}\}
     \psi_{-n,+}^{-},
\notag\\
&= \frac{1}{4mr_{0}}\left(2n\cos\theta - 1\right).
\end{align}

The resulting $z$-polarized current is particularly transparent: it is uniform along the ring and depends only on the orbital winding number and on the spin-mixing angle. The factor $\cos\theta$ shows that increasing $\xi$ progressively rotates the spin texture away from the fixed $z$ axis, thereby reducing the $z$-projected transported spin.

\begin{figure}[h!]
\centering
\includegraphics[scale=0.5]{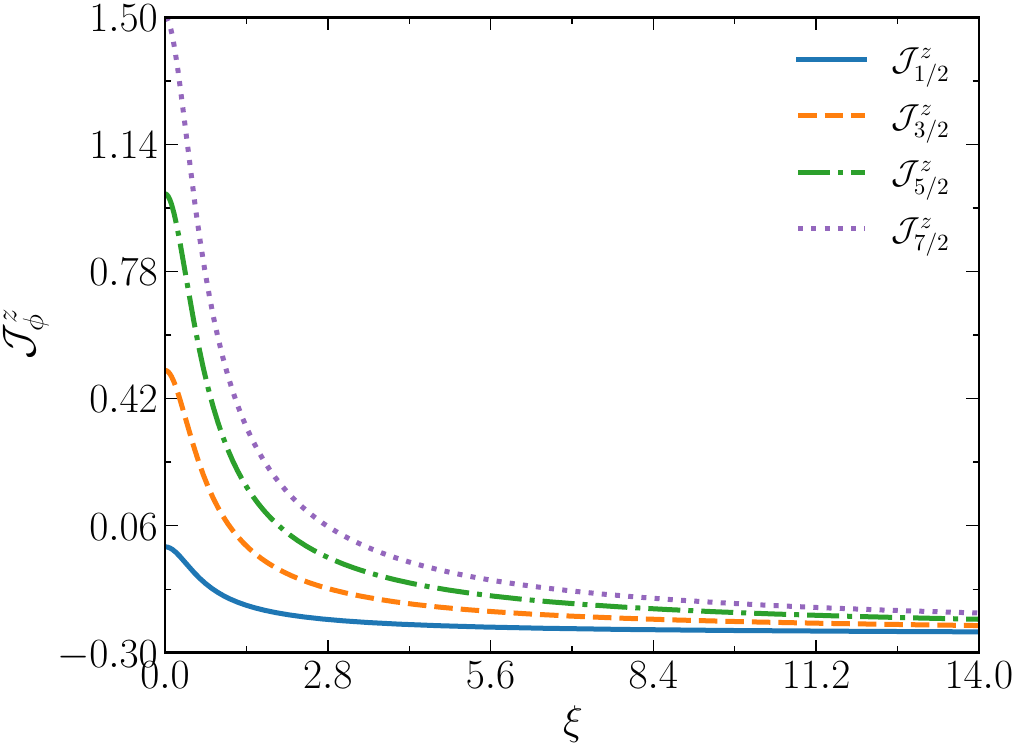}
\caption{Persistent spin current $\mathcal J_{\varphi}^{z}$ as a function of the effective coupling $\xi$ for representative orbital sectors. The current decreases as the Rashba-like interaction rotates the spin texture away from the $z$ axis, with higher-$n$ states carrying larger current because of their larger angular momentum.}
\label{fig:jz-current}
\end{figure}
As shown in Fig.~\ref{fig:jz-current}, the magnitude of $\mathcal J_{\varphi}^{z}$ is largest near weak coupling and then decreases monotonically as $\xi$ grows. This trend follows directly from $\cos\theta=(1+4\xi^{2})^{-1/2}$: the stronger the coupling, the more the spin is canted away from the $z$ direction. The different curves correspond to different orbital sectors, and larger values of $n$ start from larger currents because they represent faster winding around the ring.

This result shows that the persistent spin current is controlled by the effective mixing angle $\theta$ introduced by the Rashba-like coupling. In particular, the $\xi$ dependence enters through $\cos\theta$, while the $\xi\to0$ limit retains the residual kinematic contribution associated with the orbital quantum number $n$.

\bigskip

\noindent\textbf{Spin currents in the transverse directions.}
The $x$ and $y$ components of the spin current can also be computed following the same procedure. For each individual state they are, in general, complex quantities; however, the sum over the four degenerate states yields a real physical current. For a given mode we obtain
\begin{subequations}
\begin{align*}
\mathcal{J}_{\varphi}^{x}
 &= \frac{\mathcal{F}_{12}}{2}
    \left(2n\cos\theta - 1\right)\cos\varphi, \\
\mathcal{J}_{\varphi}^{y}
 &= \frac{\mathcal{F}_{12}}{2}
    \left(2n\cos\theta - 1\right)\sin\varphi.
\end{align*}
\end{subequations}
The explicit $\varphi$ dependence arises from the precession of the spin around the $z$ axis, which causes the polarization to rotate as the particle moves along the ring. Consequently, the sign of the spin current may oscillate as a function of the angle, even though the momentum quantum number remains fixed.
Unlike the $z$ component, the transverse currents follow the local orientation of the spin texture and are therefore explicitly angle dependent. They are phase shifted by $\pi/2$ with respect to one another, which means that the in-plane spin current vector rotates continuously as the particle propagates around the ring.

\begin{figure*}[tbhp]
\centering
\includegraphics[scale=0.55]{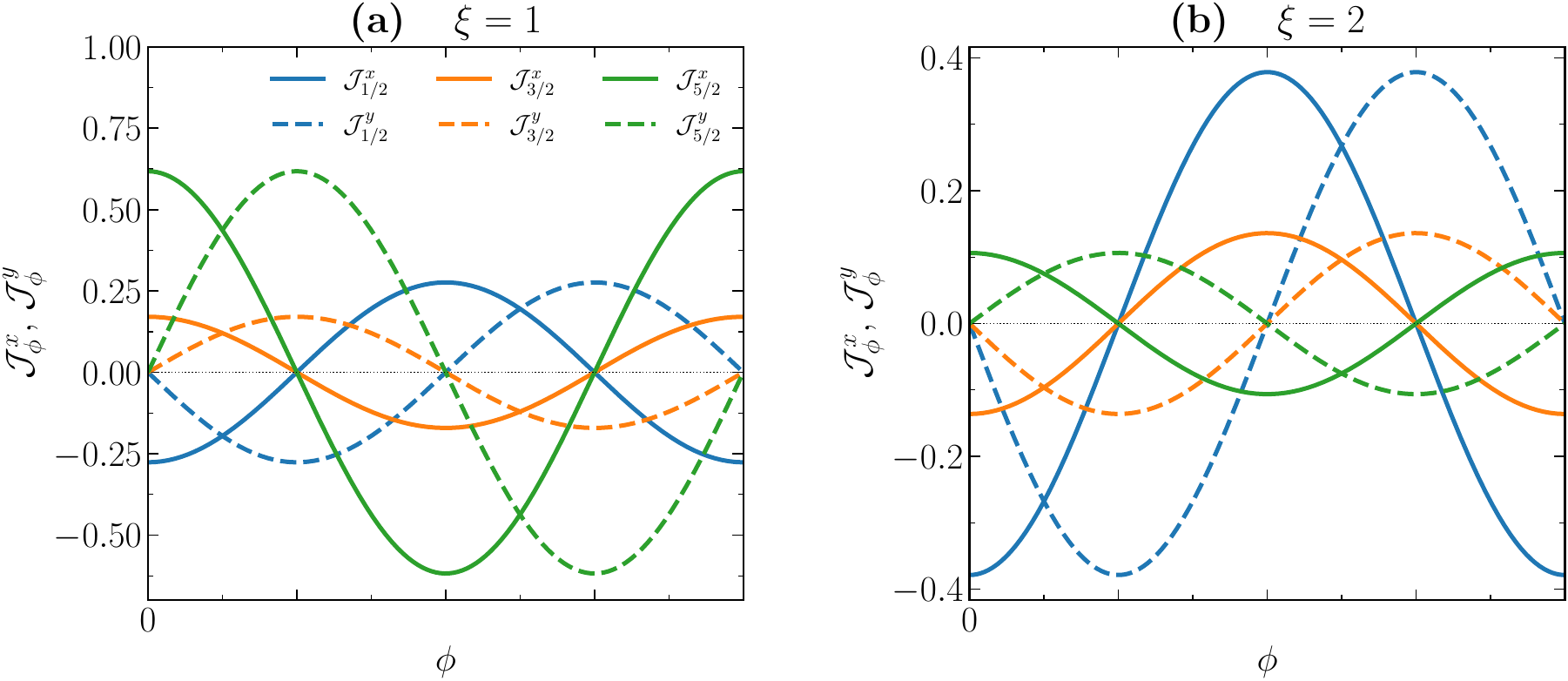}
\caption{Angular dependence of the transverse persistent spin-current components $\mathcal{J}_{\varphi}^{x}$ and $\mathcal{J}_{\varphi}^{y}$ for two representative couplings, $\xi=1$ and $\xi=2$. The sinusoidal profiles and relative phase shift illustrate the precession of the in-plane transported spin as the particle circles the ring, while the change in amplitude reflects the stronger spin locking produced by larger $\xi$.}
\label{fig:Jxy-current}
\end{figure*}
Figure~\ref{fig:Jxy-current} shows that the in-plane currents oscillate sinusoidally along the ring and that their amplitudes are enhanced when the coupling is increased from $\xi=1$ to $\xi=2$. Because $\mathcal J_{\varphi}^{x}$ and $\mathcal J_{\varphi}^{y}$ are phase shifted, they should be interpreted together as a rotating transverse current vector rather than as independent static components. In physical terms, the background-induced Rashba field locks the transported spin to the local frame of the ring and forces it to precess during the motion.

These results establish that the Rashba-like coupling produced by the field-strength derivatives naturally leads to persistent spin currents, whose magnitude and angular modulation are controlled by the effective coupling parameter $\xi$ and the geometry of the ring.

\bigskip

\noindent\textbf{Differential spin response.}
A natural measure of the response of the persistent spin current to
the effective Rashba interaction is the differential spin response,
defined as
\begin{equation}
\mathcal G_s
\equiv
\frac{\partial \mathcal J_\varphi^{\,z}}{\partial \xi}=-\frac{2 n \xi}{m r_0 (1+4\xi^2)^{\frac{3}{2}}}.
\label{eq:spin-response}
\end{equation}
The differential spin response exhibits several remarkable properties. First, it is an odd function of $\xi$, implying that reversing the
effective coupling $\xi$ reverses the spin response. In the weak
coupling regime $\xi\ll 1$, one finds a linear behavior,
\begin{equation}
\mathcal G_s \approx -\frac{2n}{m r_0}\,\xi,
\end{equation}
characteristic of a perturbative response.

In contrast, in the strong coupling regime $\xi\gg 1$, the response is
suppressed as
\begin{equation}
\mathcal G_s \sim -\frac{n}{4 m r_0}\frac{\mathrm{sign}(\xi)}{\xi^2},
\end{equation}
indicating a saturation of the spin transport.

The function reaches its maximum magnitude at
\begin{equation}
\xi_{\rm crit}=\pm\frac{1}{2\sqrt{2}},
\end{equation}
with
\begin{equation}
|\mathcal G_s|_{\max}
=
\frac{2n}{3\sqrt{3}\,m r_0}.
\end{equation}

This demonstrates the existence of an optimal coupling regime in which
spin transport is maximized. Furthermore, the response scales linearly
with the angular momentum quantum number $n$, showing that higher
orbital states contribute more efficiently to spin transport.

Finally, $\mathcal G_s$ can be interpreted as a mesoscopic spin
conductance, measuring how sensitively the persistent spin current
responds to variations of the effective coupling $\xi$, which itself
encodes the combined action of the background fields.

\bigskip

\noindent\textbf{Geometric meaning of $\mathcal G_s$.}
Since the persistent spin current along the $z$ direction is
\begin{equation*}
\mathcal J_\varphi^{z}
=
\frac{1}{4mr_0}\left(2n\cos\theta-1\right),
\end{equation*}
with
\begin{equation*}
\cos\theta=\frac{1}{\sqrt{1+4\xi^2}},
\end{equation*}
the differential spin response can be written as
\begin{equation*}
\mathcal G_s
=
\frac{\partial \mathcal J_\varphi^{z}}{\partial \xi}
=
\frac{n}{2mr_0}\frac{d(\cos\theta)}{d\xi}.
\end{equation*}
Therefore, $\mathcal G_s$ directly measures how sensitively the
spin texture is deformed by variations of the effective coupling
$\xi$. Since the same mixing angle $\theta$ also governs the
geometric phase, the differential spin response may be interpreted
as a geometric transport coefficient controlled by the parametric
evolution of the spinor texture around the ring.

From a mesoscopic perspective, $\mathcal G_s$ plays a role analogous
to a spin conductance, measuring how the persistent spin current
responds to variations of the effective coupling $\xi$, which encodes
the combined influence of the background fields. The existence of a
maximum response and its subsequent suppression at large $\xi$
highlight the nontrivial interplay between spin precession and
orbital motion in the present model.

\begin{figure}[tbh]
\centering
\includegraphics[width=\columnwidth]{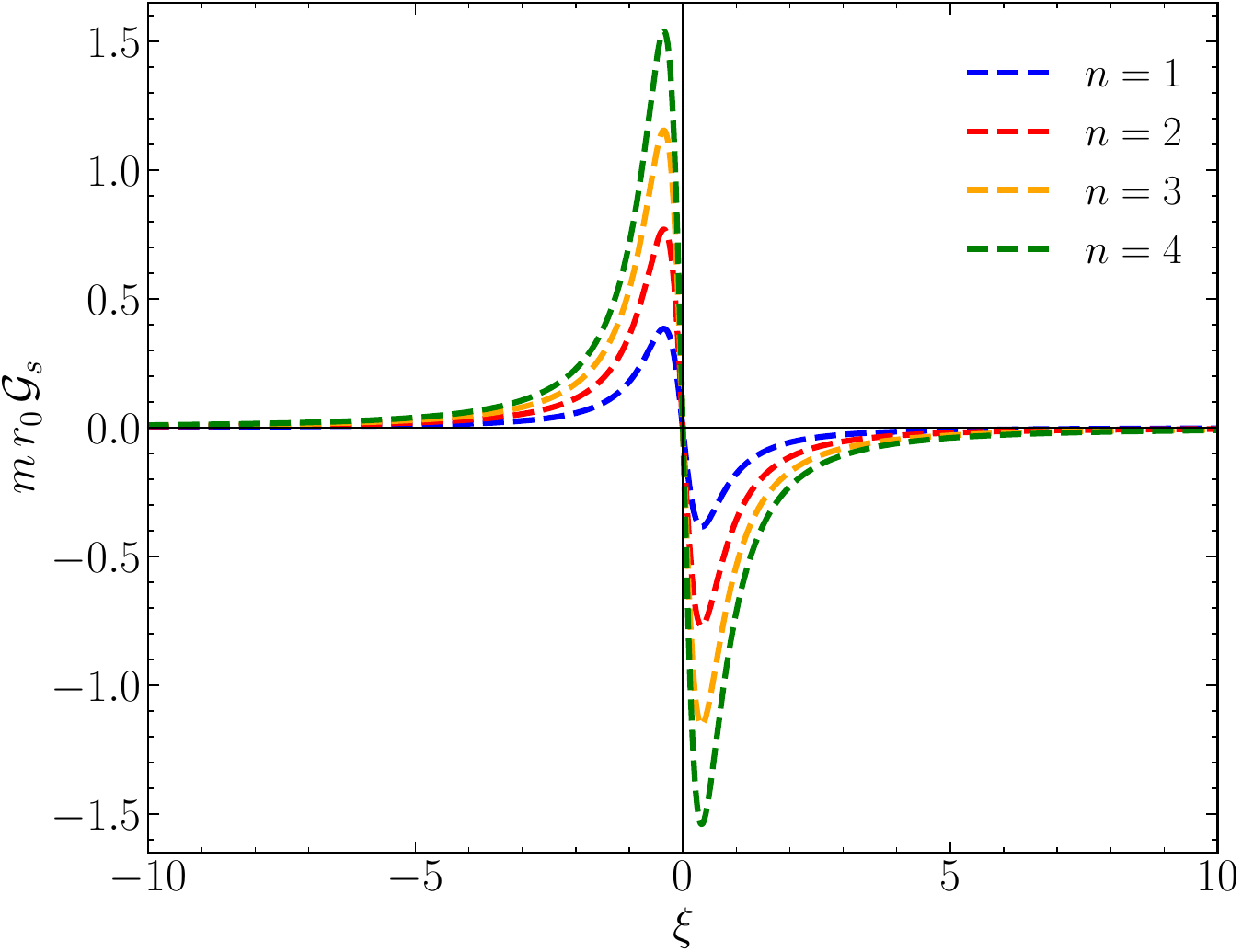}
\caption{Normalized differential spin response $m r_{0}\,\mathcal{G}_{s}$ as a function of the effective coupling $\xi$ for the representative orbital sectors $n=1,2,3,4$, computed from Eq.~\eqref{eq:spin-response}. The response is antisymmetric under $\xi\to-\xi$, vanishes at $\xi=0$, and reaches its largest magnitude at a finite coupling. The overall amplitude increases with $n$, showing that higher-winding sectors are more sensitive to variations of the Rashba-like interaction.}
\label{fig:Gs-response}
\end{figure}

Figure~\ref{fig:Gs-response} shows that the differential spin response provides a compact measure of the sensitivity of the persistent spin current to the effective coupling. Unlike the current itself, which tracks the transported spin polarization, $\mathcal{G}_{s}$ emphasizes how rapidly that transport changes as the background-induced interaction is varied. In this sense, it is natural to interpret $\mathcal{G}_{s}$ as a mesoscopic spin conductance.

Three features deserve emphasis. First, the response is odd in $\xi$, so reversing the sign of the effective coupling reverses the sign of the spin response. Second, the curves are nonmonotonic: their magnitude increases from the weak-coupling regime, reaches a maximum at the finite critical value quoted above, and then decreases for larger $|\xi|$. This identifies an optimal coupling window in which spin transport is most efficiently tunable. Third, the dependence on $n$ is purely multiplicative, so the same functional profile is preserved across orbital sectors while the response becomes stronger for larger angular momentum.

The physical origin of this behavior is geometric. As follows from the expression for $\mathcal{J}_{\varphi}^{z}$ derived in the previous subsection, the $\xi$ dependence is controlled by the spin-mixing angle $\theta$. The quantity $\mathcal{G}_{s}$ therefore measures how sensitively the local spin texture is deformed by the effective Rashba-like interaction. Since the same angle also governs the geometric phase accumulated around the ring, the response shown in Fig.~\ref{fig:Gs-response} should be viewed not only as a transport observable, but also as a geometric indicator of the parametric evolution of the spinor structure.

\section{Phenomenological bounds and experimental interpretation}
\label{sec:phenomenology}

Having developed the relativistic Dirac analysis, the nonrelativistic
Foldy--Wouthuysen reduction, and the mesoscopic ring realization, we are now
in a position to collect the phenomenological consequences of the model into a
single section. This organization is useful both conceptually and editorially:
all discussions of existing sensitivities, prospective search channels,
explicit order-of-magnitude bounds, and their physical interpretation are
gathered here only after the relevant observables have been derived in full.

Two distinct classes of references play a role in the present discussion.
First, there are works that constrain \emph{related} nonminimal operators and
therefore provide useful comparison scales for the present Lorentz-invariant
truncation, especially in EDM/MDM and SME contexts
\cite{Araujo2016,Araujo2019,KosteleckyRussell2025}. Second, there are
experimental and mesoscopic references that motivate the benchmark values used
below for electric fields, ring radii, and phase-sensitive ring observables,
such as semiconductor spin--orbit platforms and quantum-ring interferometry
\cite{Nitta1997,Morpurgo1998,Konig2006,Lorke2000,Fuhrer2001}. The estimates
presented in this section should therefore be read as literature-motivated
order-of-magnitude projections, rather than as direct extractions from a single
dedicated experiment.

\subsection{Related bounds and prospective constraints}
\label{sec:bounds}

At present, the specific Lorentz-invariant axial truncation considered in this work,
parametrized by the two effective couplings $\mathfrak g_{1}$ and $\mathfrak g_{2}$,
does not appear to possess direct experimental bounds of its own. This should not be
interpreted as an absence of phenomenological guidance, however. Rather, it reflects the
fact that the literature has mostly constrained broader or distinct classes of nonminimal
fermion--photon operators, especially in Lorentz-violating settings, while the present
model isolates a particular Lorentz-invariant sector of the dimension-six theory.

This distinction is important. In the general effective-field-theory framework, the
operators introduced in Sec.~\ref{sec:model} belong to the dimension-six fermion--photon
sector and the couplings $\mathfrak g_{1}$ and $\mathfrak g_{2}$ carry mass dimension
$-2$ \cite{PhysRevD.99.056016}. In the specific truncation adopted here, they multiply the
ordinary and dual electromagnetic tensors through the structures
$F_{\mu\nu}\gamma^\mu\gamma_5 iD^\nu$ and
$\tilde F_{\mu\nu}\gamma^\mu\gamma_5 iD^\nu$, respectively. As a consequence, the
quantities that enter observables are not $\mathfrak g_{1}$ and $\mathfrak g_{2}$ in
isolation, but rather the combinations of these couplings with the external
electromagnetic background. At the relativistic level, these combinations are naturally
encoded in the effective operator $\Gamma^\mu_{\mathrm{eff}}$ introduced above. In the
low-energy and ring limits, they reorganize into the effective parameters
$\eta_B=\mathfrak g_{1}B$, $\eta_E=\mathfrak g_{2}E$, and, in the ring geometry, the
dimensionless quantities $\xi_{1}$, $\xi_{2}$, and $\xi$.

From the viewpoint of the existing literature, the most relevant comparison is therefore
not with direct bounds on the present truncation, but with constraints on related
nonminimal sectors. For example, dimension-five CPT-even nonminimal couplings in the
Dirac equation have been constrained through the electron anomalous magnetic moment and
the electron electric dipole moment, reaching sensitivities at the level of
$10^{20}$--$10^{25}\,\mathrm{eV}^{-1}$, depending on the operator structure and on the
observable considered \cite{Araujo2016}. Likewise, dimension-six Lorentz-violating
electron--nucleon nonminimal interactions have been constrained through EDM physics at
the levels of $3.2\times10^{-13}\,\mathrm{GeV}^{-2}$ and
$1.6\times10^{-15}\,\mathrm{GeV}^{-2}$ \cite{Araujo2019}. More broadly, the nonminimal
SME data tables compile existing sensitivities for large classes of Lorentz-violating
coefficients in fermion and photon sectors of arbitrary mass dimension
\cite{KosteleckyRussell2025}. Although these results do not apply directly to the present
Lorentz-invariant truncation, they provide a useful scale of comparison for assessing the
possible reach of future searches.

In this sense, the current status of the model may be summarized as follows. The operator
class from which the present truncation descends is already phenomenologically motivated
\cite{PhysRevD.99.056016}, and neighboring sectors have been probed very deeply
\cite{Araujo2016,Araujo2019,KosteleckyRussell2025}. Nevertheless, the specific pair of
Lorentz-invariant axial couplings retained here, $\mathfrak g_{1}$ and $\mathfrak g_{2}$,
still lacks dedicated bounds. This motivates the identification of observables that are
intrinsically adapted to the structure of the theory.

Three natural routes emerge from the analyses developed in this work. The first is purely
relativistic and follows from the deformed Dirac operator itself. As shown in
Sec.~\ref{sec:relativistic}, constant electromagnetic backgrounds modify the relativistic
dispersion relation and split the propagating branches according to the orientation of the
momentum relative to the background tensor. Spectroscopic measurements of such
background-induced branch splittings would therefore constrain the effective combinations
$\eta_B=\mathfrak g_{1}B$ and $\eta_E=\mathfrak g_{2}E$ directly.

The second route is based on precision spin observables. Since related nonminimal Dirac
couplings are known to induce EDM- and MDM-like effects \cite{Araujo2016,Araujo2019},
precision spin-precession experiments performed in controlled electric or magnetic
backgrounds provide a natural arena to search for the present truncation as well. In that
context, the relevant constraint would again fall on the products of the effective
couplings with the applied fields, rather than on $\mathfrak g_{1}$ or
$\mathfrak g_{2}$ separately.

The third route is specific to the mesoscopic realization studied here. In the ring
geometry, the relativistic background couplings descend to the effective parameters
$\xi_{1}$, $\xi_{2}$, and $\xi$, which govern the spectral displacement of the branches,
the Aharonov--Anandan phase, and the persistent spin currents. Ring-based spectroscopy,
phase-sensitive interferometry, and measurements of persistent spin transport in synthetic
or condensed-matter analog platforms \cite{Nitta1997,Morpurgo1998,Konig2006,Lorke2000,Fuhrer2001}
may therefore be interpreted as direct probes of the low-energy imprint of the
relativistic theory.

For convenience, these points are summarized in Table~\ref{tab:related-bounds}. The table
is intentionally framed in terms of \emph{related bounds} and \emph{prospective search
channels}. This wording is important: the numbers quoted from the literature should be
understood as benchmarks drawn from broader or distinct operator sectors, not as direct
limits on the Lorentz-invariant truncation investigated in the present work.

\begin{table*}[t]
\centering
\caption{Representative bounds in related nonminimal sectors. The quoted values are indicative benchmarks from broader or distinct operator classes, not direct limits on the Lorentz-invariant truncation considered here. References are indicated explicitly in the last column.}
\label{tab:related-bounds}
\begin{tabular}{llll}
\hline
\textbf{Sector / class} & \textbf{Representative bound} & \textbf{Main probe / comment} & \textbf{Reference} \\
\hline
\shortstack[l]{Dimension-five nonminimal\\ fermion--photon couplings}
& $10^{-20}$--$10^{-25}\,\mathrm{eV}^{-1}$
& \shortstack[l]{Electron MDM/EDM\\ benchmark}
& \cite{Araujo2016} \\

\shortstack[l]{Dimension-six nonminimal\\ electron--nucleon couplings}
& \shortstack[l]{$3.2\times10^{-13}\,\mathrm{GeV}^{-2}$ and\\ $1.6\times10^{-15}\,\mathrm{GeV}^{-2}$}
& \shortstack[l]{Atomic EDM; distinct\\ Lorentz-violating sector}
& \cite{Araujo2019} \\

\shortstack[l]{Nonminimal SME sectors of\\ arbitrary dimension}
& Coefficient-dependent
& \shortstack[l]{Spectroscopy, cavities, clocks,\\ astrophysical and EDM-type searches}
& \cite{KosteleckyRussell2025} \\

\shortstack[l]{Present Lorentz-invariant\\ truncation}
& \shortstack[l]{No direct bound\\ presently available}
& \shortstack[l]{Relevant combinations: $\eta_B$, $\eta_E$,\\ $\xi_1$, $\xi_2$, $\xi$}
& \shortstack[l]{This work; EFT embedding in\\ \cite{PhysRevD.99.056016}} \\
\hline
\end{tabular}
\end{table*}

The main phenomenological message is therefore not that the present theory is already
bounded in a direct sense, but rather that it sits inside an operator landscape for which
high-precision searches are already known to be extremely powerful. The relativistic and
mesoscopic analyses developed here identify the combinations of couplings and backgrounds
that should be targeted in future attempts to obtain the first dedicated bounds on the
Lorentz-invariant axial truncation itself.

\subsection{Extraction of prospective bounds on $\mathfrak{g}_1$ and $\mathfrak{g}_2$}
\label{sec:bounds-extraction}

Having identified the natural observables of the theory, we now translate
experimental sensitivities into concrete order-of-magnitude estimates for
the couplings $\mathfrak{g}_1$ and $\mathfrak{g}_2$. Two complementary
regimes are considered: the relativistic sector, where the relevant
observable is the branch splitting $\Delta E = E_+-E_-$ of
Eq.~\eqref{eq:disp-magnetic-example}, and the mesoscopic sector, where the
Aharonov--Anandan phase, the persistent spin current, and the differential
spin response provide independent probes of the same effective coupling.
Throughout this subsection we work in natural units $\hbar=c=1$ and quote
all results in GeV.

The benchmark values adopted below are chosen to reflect experimentally
meaningful scales. The electric-field ranges are motivated by gate-controlled
spin--orbit systems in semiconductor heterostructures \cite{Nitta1997}, while
the ring radii and effective-mass scales are representative of mesoscopic
quantum-ring structures \cite{Lorke2000,Fuhrer2001}. Likewise, the AA-phase
benchmark is guided by phase-sensitive ring interferometry and spin-related
transport experiments \cite{Morpurgo1998,Konig2006}. The purpose here is not
to claim a unique experimental realization, but to use realistic literature-based
scales in order to assess the order of magnitude of the expected bounds.

\paragraph{Unit conversion.}
Electromagnetic fields in SI units are converted to natural units via the
Schwinger critical values. For the magnetic field,
$B_c = m_e^2/e\simeq4.41\times10^{9}\,\mathrm{T}$, so that
\begin{equation}
B\,[\mathrm{GeV}^{2}]
\simeq
1.95\times10^{-16}\,\mathrm{GeV}^{2}\times B\,[\mathrm{T}].
\label{eq:B-conversion}
\end{equation}
For the electric field, using
$E_c\simeq1.32\times10^{18}\,\mathrm{V/m}$,
\begin{equation}
E\,[\mathrm{GeV}^{2}]
\simeq
6.52\times10^{-25}\,\mathrm{GeV}^{2}\times E\,[\mathrm{V/m}].
\label{eq:E-conversion}
\end{equation}
These relations will be used systematically to express
$\eta_B=\mathfrak{g}_1 B$ and $\eta_E=\mathfrak{g}_2 E$
in terms of laboratory-accessible quantities.

\subsubsection{Relativistic sector}

\paragraph{Branch-splitting observable.}
From Eq.~\eqref{eq:disp-magnetic-example}, the energy splitting between
the two propagating branches at fixed momentum $p$ is
\begin{align}
\Delta E(p)
&=
\sqrt{m^2+p^2(1+\eta_B^2)+2m\eta_B p}\notag\\&
-
\sqrt{m^2+p^2(1+\eta_B^2)-2m\eta_B p}.
\label{eq:DeltaE-exact}
\end{align}
This expression admits transparent limiting forms in each kinematic regime.
In the \emph{nonrelativistic} limit $p\ll m$, expansion to leading order gives
\begin{equation}
\Delta E
\simeq
2\eta_B p
+\mathcal{O}\!\left(\tfrac{p^3}{m^2}\right),
\qquad p\ll m,
\label{eq:DeltaE-NR}
\end{equation}
while in the \emph{ultrarelativistic} limit $p\gg m$,
\begin{equation}
\Delta E
\simeq
2m\eta_B
+\mathcal{O}\!\left(\tfrac{m^2}{p}\right),
\qquad p\gg m.
\label{eq:DeltaE-UR}
\end{equation}
The splitting saturates to the $p$-independent value $2m\eta_B$ at high
energies. In both cases, if an experiment resolves energies with precision
$\delta E$, the condition $\Delta E < \delta E$ imposes a bound on
$\mathfrak{g}_1$.

\paragraph{Nonrelativistic scenario (low-energy electrons).}
Consider electrons with kinetic energy
$K\simeq1\,\mathrm{keV}=10^{-6}\,\mathrm{GeV}$.
The corresponding nonrelativistic momentum is not $p\simeq1\,\mathrm{keV}$,
but rather
\begin{equation}
p=\sqrt{2m_e K}
\simeq
\sqrt{2(5.11\times10^{-4})(10^{-6})}
\simeq
3.2\times10^{-5}\,\mathrm{GeV}.
\label{eq:p-kev-electron}
\end{equation}
Using Eq.~\eqref{eq:DeltaE-NR}, the condition $2\eta_B p<\delta E$ yields
\begin{equation}
\mathfrak{g}_1
<
\frac{\delta E}{2\,p\,B\,[\mathrm{GeV}^2]}.
\label{eq:bound-NR-g1}
\end{equation}
For a benchmark low-energy spectroscopic scenario with
$B=10\,\mathrm{T}$ and $\delta E=1\,\mathrm{meV}=10^{-12}\,\mathrm{GeV}$,
substituting Eq.~\eqref{eq:B-conversion}:
\begin{equation}
\mathfrak{g}_1
\lesssim
\frac{10^{-12}}{2(3.2\times10^{-5})(1.95\times10^{-15})}
\simeq
8.0\times10^{6}\,\mathrm{GeV}^{-2}.
\label{eq:bound-NR-g1-num}
\end{equation}
Although this benchmark is used here illustratively, it is physically
motivated by the general sensitivity of precision spin and spectroscopic
observables to nonminimal fermion couplings \cite{Araujo2016}.

\paragraph{Ultrarelativistic scenario (storage ring).}
For relativistic electrons with $p\simeq1\,\mathrm{GeV}$, $B=1\,\mathrm{T}$,
and relative resolution $\delta E/E\sim10^{-6}$, one has
$\delta E\sim10^{-6}\,\mathrm{GeV}$. Equation~\eqref{eq:DeltaE-UR}
then gives
\begin{equation}
\mathfrak{g}_1
<
\frac{\delta E}{2\,m_e\,B\,[\mathrm{GeV}^2]}
\simeq
5\times10^{12}\,\mathrm{GeV}^{-2}.
\label{eq:bound-UR-g1-num}
\end{equation}
The nonrelativistic scenario is therefore more constraining by roughly six
orders of magnitude, because at low momenta the splitting grows linearly
with $p$ whereas in the UR regime it saturates at $2m_e\eta_B$. This
storage-ring-like scenario should be read as a benchmark for relativistic
spectroscopy, in the same general spirit as the high-precision searches
compiled in the SME literature \cite{KosteleckyRussell2025}.

\paragraph{Electric-type branch ($\mathfrak{g}_2$).}
By the replacement $\eta_B\to\eta_E=\mathfrak{g}_2 E$, the same analysis
applies to the $\mathfrak{g}_2$ sector via Eq.~\eqref{eq:E-conversion}.
For $K=1\,\mathrm{keV}$, so that $p$ is given by
Eq.~\eqref{eq:p-kev-electron}, together with $E=10^{8}\,\mathrm{V/m}$ and
$\delta E=1\,\mathrm{meV}$:
\begin{equation}
\mathfrak{g}_2
\lesssim
\frac{10^{-12}}{2(3.2\times10^{-5})(6.52\times10^{-17})}
\simeq
2.4\times10^{8}\,\mathrm{GeV}^{-2}.
\label{eq:bound-E-g2-num}
\end{equation}
Here the benchmark electric field is motivated by the large effective fields
accessible in semiconductor spin--orbit platforms \cite{Nitta1997}.

\subsubsection{Mesoscopic sector}

In the ring geometry, the effective dimensionless coupling is
$\xi=mr_0\mathcal{F}_{12}$, which for the two models reads
\begin{align}
\xi_1 &= m^*r_0\,\mathfrak{g}_1 B,\label{eq:xi1-def}\\
\xi_2 &= m^*r_0\,\mathfrak{g}_2 E.\label{eq:xi2-def}
\end{align}
The combination $m^*r_0$ plays the role of the dimensionless lever arm
of the mesoscopic system. For a GaAs quantum ring with effective mass
$m^*=0.067\,m_e$ and radius $r_0=100\,\mathrm{nm}$, representative of
mesoscopic semiconductor-ring platforms \cite{Lorke2000,Fuhrer2001}, one finds
in natural units:
\begin{align}
m^*r_0
&=
(0.067\times 5.11\times10^{-4}\,\mathrm{GeV})\notag\\&
\times(5.07\times10^{8}\,\mathrm{GeV}^{-1})\simeq1.74\times10^{4},
\label{eq:mr0-value}
\end{align}
so that Eqs.~\eqref{eq:xi1-def}--\eqref{eq:xi2-def} become
\begin{align}
\xi_1
&\simeq
3.39\times10^{-12}\,
\mathfrak{g}_1[\mathrm{GeV}^{-2}]\,B[\mathrm{T}],
\label{eq:xi1-num}\\
\xi_2
&\simeq
1.13\times10^{-20}\,
\mathfrak{g}_2[\mathrm{GeV}^{-2}]\,E[\mathrm{V/m}].
\label{eq:xi2-num}
\end{align}
An upper bound on $\xi$ from any mesoscopic observable therefore translates
directly into a bound on $\mathfrak{g}_1$ or $\mathfrak{g}_2$ via these
relations.

\paragraph{Aharonov--Anandan phase.}
The AA phase deviates from its $\xi=0$ value by
\begin{equation}
\delta\Phi_{AA}
=
\pi s\!\left(1-\frac{1}{\sqrt{1+4\xi^2}}\right)
\simeq
2\pi s\,\xi^2,
\qquad\xi\ll1.
\label{eq:dPhiAA}
\end{equation}
If ring interferometry resolves phases to $\delta\Phi_{\min}$, the
condition $|\delta\Phi_{AA}|<\delta\Phi_{\min}$ implies
\begin{equation}
\xi
\lesssim
\sqrt{\frac{\delta\Phi_{\min}}{2\pi}}.
\label{eq:xi-AA-bound}
\end{equation}
For a benchmark phase resolution $\delta\Phi_{\min}\sim10^{-2}$, consistent
with the phase-sensitive spirit of ring-interference measurements
\cite{Morpurgo1998,Konig2006}, one finds $\xi\lesssim0.04$.
Substituting into Eqs.~\eqref{eq:xi1-num}--\eqref{eq:xi2-num} with
$B=1\,\mathrm{T}$ and $E=10^{6}\,\mathrm{V/m}$:
\begin{align}
\mathfrak{g}_1
&\lesssim
1.2\times10^{10}\,\mathrm{GeV}^{-2},
\label{eq:bound-AA-g1}\\
\mathfrak{g}_2
&\lesssim
3.5\times10^{12}\,\mathrm{GeV}^{-2}.
\label{eq:bound-AA-g2}
\end{align}

\paragraph{Persistent spin current.}
The deviation of $\mathcal{J}^z_\varphi$ from its free-ring value at
fixed $n$ is
\begin{equation}
\delta\mathcal{J}^z
\simeq
\frac{n\,\xi^2}{m^*r_0},
\qquad\xi\ll1.
\label{eq:dJz-bound}
\end{equation}
If measurements resolve spin currents to a fraction $\epsilon$ of the
zero-coupling value $(4m^*r_0)^{-1}$, then the condition
$\delta\mathcal{J}^z < \epsilon(4m^*r_0)^{-1}$ yields
\begin{equation}
\xi
\lesssim
\sqrt{\frac{\epsilon}{4n}}.
\label{eq:xi-spin-current-bound}
\end{equation}
For $n=1$ and $\epsilon=10^{-3}$, one finds
$\xi\lesssim1.58\times10^{-2}$, which is somewhat tighter than the
phase bound. Using Eqs.~\eqref{eq:xi1-num}--\eqref{eq:xi2-num} with
$B=1\,\mathrm{T}$ and $E=10^{6}\,\mathrm{V/m}$:
\begin{align}
\mathfrak{g}_1
&\lesssim
4.7\times10^{9}\,\mathrm{GeV}^{-2},
\label{eq:bound-spin-g1}\\
\mathfrak{g}_2
&\lesssim
1.4\times10^{12}\,\mathrm{GeV}^{-2}.
\label{eq:bound-spin-g2}
\end{align}
The geometric and transport context for these benchmarks is again that of
semiconductor quantum rings and mesoscopic spin transport
\cite{Lorke2000,Fuhrer2001,Morpurgo1998,Konig2006}.

\paragraph{Differential spin response.}
The differential spin response $\mathcal{G}_s$,
Eq.~\eqref{eq:spin-response}, for $\xi\ll1$ reduces to
\begin{equation}
\mathcal{G}_s \simeq -\frac{2n}{m^*r_0}\,\xi.
\end{equation}
Hence, a measurement with absolute uncertainty $\delta\mathcal{G}_s$
would constrain
\begin{equation}
\xi
\lesssim
\frac{m^*r_0\,}{2n}|\delta\mathcal{G}_s|.
\label{eq:xi-Gs-bound}
\end{equation}
This observable is therefore complementary to the AA phase and to the spin
current: while the latter probe $\xi^2$ in the weak-coupling regime, the
differential spin response is itself linear in $\xi$ and may thus provide a
first-derivative probe of small departures from the uncoupled limit.

\subsubsection{Summary and comparison}

The bounds derived above are collected in Table~\ref{tab:bounds-extracted}.
Several observations are in order.

First, the nonrelativistic spectroscopic scenario consistently outperforms
the ultrarelativistic one for $\mathfrak{g}_1$, because the splitting
$\Delta E\simeq2\eta_Bp$ grows with $p$ in the NR regime, whereas in
the UR regime it saturates at $2m_e\eta_B$.

Second, the $\mathfrak{g}_2$ bounds benefit from the large electric fields
achievable in semiconductor heterostructures \cite{Nitta1997},
partially compensating the smaller conversion factor in
Eq.~\eqref{eq:E-conversion}.

Third, the corrected estimated bounds ($\sim10^{6}$--$10^{12}\,\mathrm{GeV}^{-2}$)
are numerically much weaker than the literature benchmarks in
Table~\ref{tab:related-bounds} ($\sim10^{-13}$--$10^{-15}\,\mathrm{GeV}^{-2}$).
This contrast is physically meaningful rather than discouraging: the
literature constraints target \emph{Lorentz-violating} operators, whose
coefficients encode a preferred frame. The couplings $\mathfrak{g}_1$ and
$\mathfrak{g}_2$ are \emph{Lorentz-invariant} and carry no such
suppression; the estimates in Table~\ref{tab:bounds-extracted} should
therefore be viewed as the first dedicated order-of-magnitude bounds on
this specific truncation. Improving the magnetic field, ring radius, or
current sensitivity by one to two orders of magnitude would push the
mesoscopic bounds appreciably downward.

\begin{table*}[t]
%\footnotesize
\scriptsize
\centering
\caption{Order-of-magnitude prospective bounds on $\mathfrak{g}_1$ and
$\mathfrak{g}_2$ extracted from the observables of the present model in the
relativistic and mesoscopic regimes. All estimates assume
laboratory-accessible experimental conditions and should be interpreted as
order-of-magnitude projections; dedicated experiments targeting these
observables have not yet been performed. The last column indicates the main
literature motivation for the benchmark values adopted in each row.}
\label{tab:bounds-extracted}
\begin{tabular}{llllll}
\hline
\textbf{Regime} & \textbf{Observable} & \textbf{Coupling} &
\textbf{Bound (GeV$^{-2}$)} & \textbf{Scenario / assumptions} &
\textbf{Reference / motivation} \\
\hline
Relativistic (NR)
& $\Delta E\simeq2\eta_B p$
& $\mathfrak{g}_1$
& $\lesssim 8.0\times10^{6}$
& $K=1\,\mathrm{keV}$, $B=10\,\mathrm{T}$, $\delta E=1\,\mathrm{meV}$
& \shortstack[l]{Benchmark low-energy spectroscopy;\\ related precision-spin sensitivity \cite{Araujo2016}} \\

Relativistic (UR)
& $\Delta E\simeq2m\eta_B$
& $\mathfrak{g}_1$
& $\lesssim 5\times10^{12}$
& $B=1\,\mathrm{T}$, $p=1\,\mathrm{GeV}$, $\delta E/E=10^{-6}$
& \shortstack[l]{Benchmark relativistic spectroscopy /\\ storage-ring-type scenario; cf. \cite{KosteleckyRussell2025}} \\

Relativistic (NR)
& $\Delta E\simeq2\eta_E p$
& $\mathfrak{g}_2$
& $\lesssim 2.4\times10^{8}$
& $K=1\,\mathrm{keV}$, $E=10^{8}\,\mathrm{V/m}$, $\delta E=1\,\mathrm{meV}$
& \shortstack[l]{Electric-field scale motivated by\\ semiconductor tunability \cite{Nitta1997}} \\

Mesoscopic
& AA phase $\delta\Phi_{AA}$
& $\mathfrak{g}_1$
& $\lesssim 1.2\times10^{10}$
& $B=1\,\mathrm{T}$, $r_0=100\,\mathrm{nm}$, $\delta\Phi=10^{-2}$
& \shortstack[l]{Ring interferometry and SO phases\\ \cite{Morpurgo1998,Konig2006}; geometry \cite{Lorke2000,Fuhrer2001}} \\

Mesoscopic
& AA phase $\delta\Phi_{AA}$
& $\mathfrak{g}_2$
& $\lesssim 3.5\times10^{12}$
& $E=10^{6}\,\mathrm{V/m}$, $r_0=100\,\mathrm{nm}$, $\delta\Phi=10^{-2}$
& \shortstack[l]{Phase-sensitive rings \cite{Morpurgo1998,Konig2006};\\ electric-field control \cite{Nitta1997}} \\

Mesoscopic
& Spin current $\mathcal{J}^z_\varphi$
& $\mathfrak{g}_1$
& $\lesssim 4.7\times10^{9}$
& $B=1\,\mathrm{T}$, $r_0=100\,\mathrm{nm}$, $\epsilon=10^{-3}$, $n=1$
& \shortstack[l]{Quantum-ring transport scales\\ \cite{Lorke2000,Fuhrer2001}} \\

Mesoscopic
& Spin current $\mathcal{J}^z_\varphi$
& $\mathfrak{g}_2$
& $\lesssim 1.4\times10^{12}$
& $E=10^{6}\,\mathrm{V/m}$, $r_0=100\,\mathrm{nm}$, $\epsilon=10^{-3}$, $n=1$
& \shortstack[l]{Quantum-ring transport context\\ \cite{Lorke2000,Fuhrer2001}; field control \cite{Nitta1997}} \\
\hline
\end{tabular}
\end{table*}

\subsection{Physical interpretation of the magnitude of the bounds}
\label{sec:bounds-interpretation}

At first sight, the order-of-magnitude bounds obtained for the couplings
$\mathfrak g_1$ and $\mathfrak g_2$ may look surprisingly large, especially
when compared with the much stronger limits usually quoted in the literature
for Lorentz-violating nonminimal sectors \cite{Araujo2016,Araujo2019,KosteleckyRussell2025}.
This, however, does \emph{not} signal any inconsistency of the present analysis.
Rather, it reflects a basic feature of the Lorentz-invariant dimension-six
truncation considered here: physical observables do not depend on
$\mathfrak g_1$ and $\mathfrak g_2$ in isolation, but only through the
combinations of these couplings with the external electromagnetic fields.

More precisely, in the relativistic sector the relevant dimensionless
parameters are
\begin{equation}
\eta_B = \mathfrak g_1 B,
\qquad
\eta_E = \mathfrak g_2 E,
\end{equation}
whereas in the mesoscopic ring realization the corresponding control parameter
is
\begin{equation}
\xi \sim m^* r_0\, \mathfrak g\,F,
\end{equation}
with $F$ standing schematically for the effective electromagnetic background.
Therefore, a numerically large upper bound on $\mathfrak g_1$ or
$\mathfrak g_2$ simply means that the present experimental scenario is only
weakly sensitive to the products $\mathfrak g_1 B$, $\mathfrak g_2 E$, or
$\xi$, not that the induced physical effect is large.

This point becomes especially transparent in natural units. Since
\begin{align}
&1\,\mathrm{T} \simeq 1.95\times10^{-16}\,\mathrm{GeV}^2,
\\
&1\,\mathrm{V/m} \simeq 6.52\times10^{-25}\,\mathrm{GeV}^2,
\end{align}
even laboratory fields that are large in SI units are extremely small in
GeV$^2$. For instance, taking $\mathfrak g_1\sim10^{10}\,\mathrm{GeV}^{-2}$
and $B=1\,\mathrm{T}$, one obtains
\begin{equation}
\eta_B=\mathfrak g_1 B
\sim 10^{10}\times1.95\times10^{-16}
\sim2\times10^{-6},
\end{equation}
which is a tiny dimensionless deformation. Thus, a large numerical bound on
$\mathfrak g_1$ is perfectly compatible with a very small physical effect.

This observation is also natural from the effective-field-theory viewpoint.
Since the couplings have mass dimension $-2$, one may write schematically
\begin{equation}
\mathfrak g_i \sim \frac{1}{\Lambda_i^2},
\end{equation}
so that a bound such as
$\mathfrak g_i \lesssim 10^{10}\,\mathrm{GeV}^{-2}$ corresponds only to
\begin{equation}
\Lambda_i \gtrsim |\mathfrak g_i|^{-1/2}
\sim 10^{-5}\,\mathrm{GeV}\sim 10\,\mathrm{keV}.
\end{equation}
Therefore, the present bounds should be interpreted as
\emph{weak first prospective limits} on a Lorentz-invariant dimension-six
operator, rather than as precision constraints in the usual EFT sense.

The relativistic and mesoscopic sectors are fully consistent with one
another in this respect. In the relativistic regime, the branch splitting is
controlled by the field-dressed combinations $\eta_B=\mathfrak g_1 B$ and
$\eta_E=\mathfrak g_2 E$, not by $\mathfrak g_1$ and $\mathfrak g_2$
separately. In the nonrelativistic limit,
\begin{equation}
\Delta E_{\rm NR}\simeq2\eta_B p,
\end{equation}
whereas in the ultrarelativistic limit
\begin{equation}
\Delta E_{\rm UR}\simeq2m\eta_B.
\end{equation}
This explains why the ultrarelativistic scenario is less constraining:
the splitting ceases to grow with momentum and saturates at a value
controlled by the electron mass. In the mesoscopic regime, the same logic
applies through the parameter $\xi$, which controls the AA phase, the spin
current, and the differential response. Large numerical bounds on
$\mathfrak g_1$ and $\mathfrak g_2$ are therefore the natural signal of a
dimension-six Lorentz-invariant operator being tested only through
field-dressed combinations in laboratory-scale electromagnetic backgrounds.

The main physical message is thus not that the model predicts large effects,
but rather that the present experimental scenarios are only weakly sensitive
to this operator class. The large bounds obtained for $\mathfrak g_1$ and
$\mathfrak g_2$ should be interpreted as \emph{weak prospective constraints}
on a Lorentz-invariant dimension-six truncation. This is entirely compatible
with the fact that the observable deformations of the spectrum, geometric
phase, and persistent spin currents are all perturbatively small.

It is also worth stressing that this situation differs conceptually from the
one found in Lorentz-violating nonminimal sectors. In those cases, the
coefficients are often constrained much more strongly because the associated
observables probe preferred-frame effects with extremely high precision
\cite{Araujo2016,Araujo2019,KosteleckyRussell2025}. Here, by contrast, the
couplings are Lorentz-invariant, and the only way to test them is through the
background-dressed products $\mathfrak g_1 B$, $\mathfrak g_2 E$, and $\xi$.
The resulting sensitivities are therefore naturally much weaker.

Finally, one may summarize the present status as follows: the largeness of the
bounds is physically meaningful and internally consistent, but it also shows
that the present estimates should be viewed as first order-of-magnitude
projections rather than precision limits. Their real value is that they
identify which experimental channels --- low-energy spectroscopy, phase
interferometry, and persistent spin transport --- are the most promising ones
for obtaining the first dedicated bounds on this Lorentz-invariant axial
truncation.

\section{Conclusion}
 
In this work we have investigated a class of nonminimal derivative couplings
between fermions and electromagnetic tensor backgrounds, characterized by two
Lorentz-invariant dimensionful couplings $\mathfrak{g}_1$ and $\mathfrak{g}_2$.
The analysis proceeded across three complementary levels --- the relativistic
Dirac theory, the Foldy--Wouthuysen nonrelativistic reduction, and the
mesoscopic quantum ring --- and the coherence of the results across these levels
constitutes one of the central messages of the paper.
 
Already at the Dirac level, before any nonrelativistic approximation is
performed, the deformed operator $\Gamma^\mu_{\mathrm{eff}}$ splits the
relativistic mass shell into two distinct branches $E_\pm$, whose separation
grows as $\Delta E\simeq2\eta_Bp$ at low momenta and saturates at
$\Delta E\simeq2m\eta_B$ in the ultrarelativistic regime, as illustrated in
Fig.~\ref{fig:dispersion-magnetic}. The canonical analysis reveals that
admissible background configurations are precisely those for which
$\Gamma^0_{\mathrm{eff}}$ is invertible, so that a consistent one-particle
Hamiltonian exists; the effective bilinear current
$J^\mu_{\mathrm{eff}}=\bar\psi\,\Gamma^\mu_{\mathrm{eff}}\psi$, conserved for
constant backgrounds, then provides the appropriate relativistic probability
density. These relativistic features are not secondary details: they identify
the branch splitting as a genuine kinematic signature of the model, already
present before any reduction to the ring geometry is performed.
 
The Foldy--Wouthuysen transformation converts this relativistic content into
a transparent low-energy structure. Both the $\mathfrak{g}_1$ and $\mathfrak{g}_2$
sectors generate effective Hamiltonians of the form
$\boldsymbol{\mathcal{F}}\cdot(\boldsymbol{p}\times\boldsymbol{\sigma})$,
showing that magnetic as well as electric background fields can induce Rashba-type
interactions. This is a genuinely novel feature of the present framework: in
standard condensed-matter realizations the Rashba coupling is exclusively
driven by structural electric fields~\cite{Bychkov1984}, whereas here both
$\mathbf{B}$ and $\mathbf{E}$ may play equivalent roles depending on which
operator sector is active. When restricted to a one-dimensional quantum ring,
the effective Hamiltonian takes the exact form
$(i\partial_\varphi+\xi\sigma_\rho)^2/(2mr_0^2)-\xi^2/(2mr_0^2)$, where
the single dimensionless parameter $\xi=mr_0\mathcal{F}_{12}$ unifies, in one
stroke, the spectral splitting, the geometric phase, and the persistent spin
current of the ring.
 
From the exact solution of the ring problem we derived analytical expressions
for the spectrum, eigenspinors, Aharonov--Anandan phases, and persistent spin
currents. The AA phase
$\Phi_{AA}^{(\lambda,s)}=-2\lambda\pi(n-\frac{\lambda s}{2}\cos\theta)$
encodes the geometric structure of the spinor bundle over the ring through the
mixing angle $\theta=\arctan(2\xi)$, which measures the deformation of the
local spin frame induced by the nonminimal coupling. The persistent spin
current in the $z$ direction, $\mathcal{J}_\varphi^z=(4mr_0)^{-1}(2n\cos\theta-1)$,
is uniform along the ring and decreases monotonically as $\xi$ grows
(Fig.~\ref{fig:jz-current}), while the transverse components
$\mathcal{J}_\varphi^{x,y}$ oscillate sinusoidally with the ring angle,
forming a rotating in-plane current vector locked to the local spin frame
(Fig.~\ref{fig:Jxy-current}). The differential spin response
$\mathcal{G}_s=\partial\mathcal{J}_\varphi^z/\partial\xi$, which plays the
role of a geometric transport coefficient, reaches its maximum magnitude at
the finite value $\xi_{\rm crit}=1/(2\sqrt{2})$ with
$|\mathcal{G}_s|_{\max}=2n/(3\sqrt{3}mr_0)$, signaling an optimal coupling
regime for spin manipulation. This feature, together with the linear scaling
with the orbital quantum number $n$, may be of direct relevance for spintronic
applications~\cite{Zutic2004,Manchon2015} where efficient and tunable spin
control is a central requirement.
 
A central new contribution of this work is the first systematic extraction
of order-of-magnitude bounds on the two Lorentz-invariant couplings from
experimentally accessible scenarios, collected in Table~\ref{tab:bounds-extracted}.
In the relativistic sector, the nonrelativistic kinematic regime turns out to
be the most sensitive: for electrons with kinetic energy
$K\sim1\,\mathrm{keV}$, magnetic field $B=10\,\mathrm{T}$, and spectroscopic
resolution $\delta E=1\,\mathrm{meV}$, the branch splitting implies
$\mathfrak{g}_1\lesssim8\times10^6\,\mathrm{GeV}^{-2}$, while the analogous
electric branch gives $\mathfrak{g}_2\lesssim2.4\times10^8\,\mathrm{GeV}^{-2}$.
These estimates outperform the ultrarelativistic storage-ring scenario for
$\mathfrak g_1$ by several orders of magnitude, because the low-energy
splitting grows linearly with momentum whereas the high-energy one saturates at
$2m_e\eta_B$. In the mesoscopic sector, phase-sensitive interferometry and
spin-current measurements in GaAs quantum rings give bounds in the range
$10^{10}$--$10^{12}\,\mathrm{GeV}^{-2}$, depending on whether the relevant
observable is the Aharonov--Anandan phase or the persistent spin current.
Although these estimates are numerically much weaker than the limits available
for related Lorentz-violating sectors quoted in Table~\ref{tab:related-bounds},
the comparison is physically misleading: the couplings $\mathfrak{g}_1$ and
$\mathfrak{g}_2$ are Lorentz-invariant and enter observables only through the
field-dressed combinations $\eta_B=\mathfrak g_1 B$, $\eta_E=\mathfrak g_2 E$,
and $\xi\sim m^*r_0\,\mathfrak g\,F$. Since laboratory fields are extremely
small in natural units, large numerical bounds on $\mathfrak g_1$ and
$\mathfrak g_2$ are fully consistent with perturbatively small physical
effects. The entries of Table~\ref{tab:bounds-extracted} should therefore be
viewed as the first dedicated order-of-magnitude constraints on this specific
operator truncation rather than as precision EFT limits.

Altogether, these findings reveal that nonminimal couplings involving tensor
fields can generate a phenomenology remarkably similar to Rashba systems,
while retaining a fully field-theoretic foundation. This opens new paths for
connecting high-energy theoretical structures, such as antisymmetric-tensor
backgrounds, axion-like fields~\cite{Wilczek1987}, or noncommutative
corrections, to condensed-matter analogues exhibiting spin--orbit physics,
geometric phases, and mesoscopic transport. The framework developed here sets
the stage for future investigations encompassing time-dependent backgrounds,
disorder, collective behavior in arrays of rings, and possible topological
extensions. On the experimental side, engineered synthetic gauge fields in
ultracold atomic systems~\cite{Dalibard2011,Goldman2014} and topological
photonic lattices~\cite{Ozawa2019} offer platforms where $\xi$ can be
directly tuned, making the effective Hamiltonians derived here accessible in
controlled laboratory settings. These directions may broaden the scope of
applications and shed further light on the deep interplay between relativistic
field-theory structures, geometric phases, spin dynamics, and thermal
transport in low-dimensional quantum systems.

\section*{Acknowledgments}
This work was partially supported by the Brazilian agencies CAPES, CNPq, FAPEMA and FAPESB. EOS acknowledges the support from grants CNPq/306308/2022-3,
FAPEMA/UNIVERSAL-06395/22, and CAPES/Code 001. J.A.A.S.R acknowledges partial financial support from UESB through Grant AuxPPI (Edital No. 267/2024), as well as from FAPESB--CNPq/Produtividade under Grant No. 12243/2025 (TOB-BOL2798/2025).

\end{document}